\documentclass[%
 reprint,
superscriptaddress,
 amsmath,amssymb,
 aps,
floatfix,
]{revtex4-2}

\usepackage{graphicx}
\usepackage{dcolumn}
\usepackage{bm}
\usepackage{graphicx}
\usepackage{caption}
\usepackage{subcaption}

\usepackage{xcolor}
\usepackage{xfrac}
\usepackage{float}

\usepackage{mathtools}

\usepackage{comment}

\begin{document}

\preprint{APS/123-QED}

\title{Dynamic wetting experiments with nitrogen \\in a quasi-capillary tube}

\author{Domenico Fiorini}
\email{domenico.fiorini@vki.ac.be}
\affiliation{%
	von Karman Institute for Fluid Dynamics, Waterloosesteenweg 72, Sint-Genesius-Rode, Belgium
}%
\affiliation{%
	KU Leuven, Dept. of Materials Engineering, Leuven 3001, Belgium
}%
\author{Alessia Simonini}
\affiliation{%
	von Karman Institute for Fluid Dynamics, Waterloosesteenweg 72, Sint-Genesius-Rode, Belgium
}%
\author{Johan Steelant}
\affiliation{%
	ESTEC-ESA, Keplerlaan 1, Noordwijk, The Netherlands
}%
\author{David Seveno}
	\affiliation{%
	KU Leuven, Dept. of Materials Engineering, Leuven 3001, Belgium
}%
\author{Miguel Alfonso Mendez}
\affiliation{%
	von Karman Institute for Fluid Dynamics, Waterloosesteenweg 72, Sint-Genesius-Rode, Belgium
}%

\date{\today}

\begin{abstract}
This work investigates the wetting dynamics of cryogenic fluids in inertia-dominated conditions. We experimentally characterized an oscillating gas-liquid interface of liquid nitrogen in a partially filled U-shaped quartz tube. The experiments were carried out in controlled cryogenic conditions, with interface oscillations produced by releasing the liquid column from an unbalanced position and having nitrogen vapor as the only ullage gas. During the experiments, the interface shape was tracked via image processing and used to fit a model from which the contact angle could be accurately determined. The results show that the dynamic contact angle evolution in advancing conditions is linearly linked to the Capillary number, with a slope depending on whether the interface moves over a dry or a pre-wet surface. However, the contact angle remains close to the one at equilibrium in receding conditions. To analyze the relation between contact angle and interface dynamics, we define an equivalent contact angle as the one that would make a spherical interface produce the same capillary pressure drop as the actual interface shape. The evolution of this equivalent contact angle proved to be independent of the evolution of the actual one, suggesting that the interface shape is not influenced by it. Finally, a theoretical analysis of the interface motion using a simplified model shows that viscous forces dominate the damping of the interface for small tube sizes, while gravity and inertial forces dominate the oscillating dynamics of the liquid column for larger tubes.
\end{abstract}

\maketitle


\section{Introduction}



Predicting the capillary-driven motion of a liquid is essential in developing propellant management devices \cite{jaekle1997propellant,hartwig2017propellant,levine2015surface,white2019capillary} and heat transfer systems for space applications \cite{Wanison2020,Foster1973,NASA}. Moreover, capillary forces play a major role in the dynamics of cryogenic propellants in partially filled tanks \cite{kulev2010drop,kulev2014interface,schmitt2015free} in microgravity. Modeling the contact line and contact angle dynamics is essential for simulating sloshing motion and the evaporation rate in propellant tanks \cite{schmitt2015free,bellur2016visualization}. Both phenomena need to be accurately controlled to ensure that no gas is fed to the thrusters, to limit undesired loads on the tank walls and perturbations of the spacecraft stability \cite{marques2022experimental}.  

Developing engineering models for these applications requires experimental data on the dynamic wetting of cryogenic fluids such as liquid Oxygen, Hydrogen, or Methane. 
These fluids have low surface tension, low viscosity, near-zero contact angle, and high volatility. Experimental data on these liquids are particularly scarce, as most of the literature has focused on fluids with the opposite properties (particularly high viscosity and surface tension \cite{hoffman1975study,voinov1976hydrodynamics,blake1969kinetics,tanner1979spreading}). Besides challenging cryogenic temperatures, experiments on the dynamic wetting of cryogenic liquids require complex experimental setups to cope with the high volatility, promoting evaporation/condensation \cite{bellur2016visualization} and the possible occurrence of film boiling \cite{Hsu1972,Chen2020}.

Within the framework of microgravity experiments, 
\citet{friese2019liquid} investigated the axial sloshing of cryogenic hydrogen and the corresponding contact line in microgravity conditions and super-heated walls. The authors observed that the axial sloshing was only affected when the apparent wall contact line receded and the liquid film at the wall dried out. However, no visualization of the contact line was reported, and the authors suggest performing more experimental and theoretical investigations with smaller test cases to understand the contact line dynamics.

The current work presents the experimental characterization of dynamic wetting of liquid nitrogen (LN2) in cryogenic conditions. We measured the evolution of the dynamic contact angle and the gas-liquid interface within a wide range of contact-line velocity and acceleration and analyzed the relative impact of inertia, capillary and viscous forces. The experiments were carried out on a U-shaped quartz tube in which liquid oscillations were produced by releasing the liquid column from an initially unbalanced configuration. The wetting dynamic in this configuration is similar to what is observed in the forced liquid plug flows \cite{Mamba2018,Aussillous2000,Klaseboer2014} and has also been investigated by 
\citet{weislogel1996steady},
\citet{dollet2020transition}, and \citet{fiorini2022effect}. More specifically, \citet{weislogel1996steady} considered silicone oil in microgravity conditions with different surface coating on the two sides to produce a capillary-driven flow while \citet{dollet2020transition}'s experiments were carried out in normal gravity conditions using pure water and ethanol to analyze the impact of wetting hysteresis on the oscillating dynamics. \citet{fiorini2022effect} focused on the impact of inertia in the dynamic wetting of HFE7200, a well-known cryogenic model fluid, and demineralized water.

The rest of the paper is organized as follows. Section \ref{sec:modeling} analyze a simple engineering model of the interface dynamic along with the relevant dimensionless number and scaling considerations. Section \ref{sec:methodology} describes the cryogenic facility where the experiments were conducted, along with the experimental procedure, interface tracking, and contact angle measurement.  Section \ref{sec:results} presents the results for various initial heights and a discussion of the relative importance of the forces governing the flow in this configuration. Conclusions and outlooks for future works are collected in section \ref{sec:conclusions}.

\section{Modeling and Scaling Considerations}\label{sec:modeling}

\begin{figure}[h]
\centering
\includegraphics[width=6.5cm,clip]{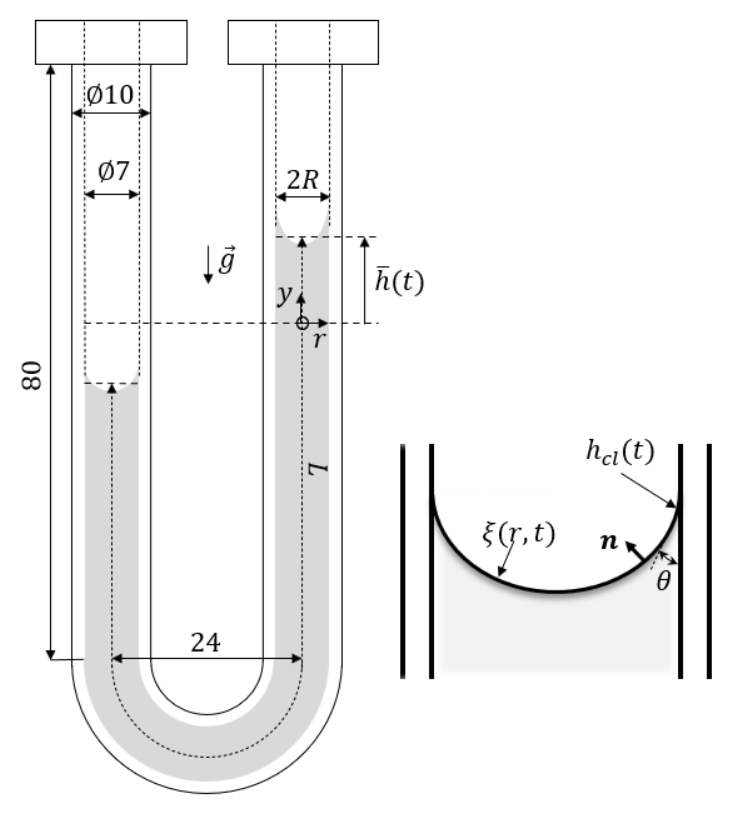}
\caption{Schematic of U-tube with dimensions in millimeters. The relevant experimental variables are also shown on the right side of the tube and in the close-up view of the gas-liquid interface.}
\label{fig:UtubeDimensions}       
\end{figure}

A schematic of the U-tube test case is provided in Figure \ref{fig:UtubeDimensions} along with the main dimensions and relevant definitions. The U-tube is made of transparent quartz. It has a constant (internal) radius $R=3.5$ mm and is filled with liquid nitrogen to have a liquid column of axial length $L=99\pm2$ mm. A meniscus is formed on each side of the tube. We denote as $h(r,t)$ the interface height with respect to the equilibrium position (i.e. $h\rightarrow 0$ as $t\rightarrow \infty$), with $r$ the radial coordinate. We denote with $\bar{h}(t)$ the average column height defined as
\begin{equation}
    \bar{h}\left(t\right)=\frac{1}{\pi R^2}\int_{0}^{R}{2\pi\ h\left(r,t\right)rdr}\label{eq:h_average}\,.
\end{equation}
For later convenience, we define as $\xi(r,t)$ the interface position with respect to $h(0,t)$, i.e $\xi(r,t)=h(r,t)-h(0,t)$ at each time step (see Figure \ref{fig:UtubeDimensions}). The liquid properties of interest are the liquid dynamic viscosity $\mu$ and density $\rho$ and the gas-liquid interface's surface tension $\sigma$.

The experiments begin by releasing the liquid column from an initial height of $\overline{h}(0)=\pm18.4 \pm 1.5$mm. This is achieved by initially pressurizing one side of the tube.

We consider the modeling of the problem from two different length scales: (1) the interface scale, which controls the interface dynamics and is concerned with forces acting close to it and (2) a tube scale which controls the liquid column's dynamics and is concerned with forces acting all along the tube. We assume that scale (1) controls the shape of the interface, which in turns play a role in (2) through the capillary pressure drop due to the interface curvature.

The treatment of scale (2) can be made in terms of integral balance of forces in the liquid column. The integral balance for the column gives
\begin{equation}
    \ddot{\bar{h}}(t)=\underbracket[0.8pt]{-8 C_f\frac{\mu}{\rho R^2}\dot{\bar{h}}(t)}_\text{\clap{viscous resistance~}}\ \ -\underbracket[0.8pt]{\frac{2g}{L}\ \bar{h}(t)}_\text{\clap{gravity~}}-\underbracket[0.8pt]{\frac{2\sigma}{\rho R^2L}\left(K_A(t)-K_B(t)\right)}_\text{\clap{capillary resistance~}}\label{eq:utubedimensional}\,,
\end{equation} where a parabolic velocity profile is assumed in the stream-wise direction when computing the wall shear and $C_f$ is an empirical term to correct this assumption, to handle the loss of parabolicity due to inertia and the viscous losses due to the tube's curvature \cite{dollet2020transition}. The dot denotes differentiation in time. The terms $K_A$ and $K_B$ are linked to the pressure drop produced at the two interfaces (distinguished with A and B) and accounts for the interface curvature. 

Under the assumption that capillary forces dominates over elongational viscosity, this term can be written as 
\begin{equation}
    K(t)=\int_{0}^{R}{\nabla\cdot\mathbf{n}(\xi(r,t))} r dr\,\label{eq:Kterm}
\end{equation} where $\mathbf{n}$ is the normal vector to the interface (see Figure \ref{fig:UtubeDimensions}), and $\nabla \cdot ()$ is the divergence operator. For capillary tubes, and particularly at the limit $R/l_c\ll1 $, with $l_c=\sqrt{\sigma/(\rho g)}$ the capillary length, or at the limit of Capillary number $Ca=\mu u_{c}/\sigma\ll1$, with $u_c$ the contact-line velocity, the interface shape is a spherical cap and thus one recovers $K=R \cos(\theta_D)$ with $\theta_D$ the (dynamic) contact angle at the wall. If one of the two sides has a flat interface (non-wetting conditions) or if one considers a straight tube plunged into a bath, equation \eqref{eq:utubedimensional} is essentially a variant of Lucas–Washburn-s equation \cite{Washburn1921,quere1997inertial,zhong2019analytic,stange2003capillary}, whose steady state solution is the well known Jurin's law \cite{Liu2018}. In quasi capillary tubes (i.e. $R/l_c> 1$), in case of $Ca> 10^{-3}$, or in the presence of large accelerations, the meniscus shape departs from that of a spherical cap. In the modeling framework considered in this work, this results from the interface scale.

\subsection{Interface modeling}
The empirical model of the meniscus interface proposed by \citet{fiorini2022effect} computes the meniscus profile $\xi(r,t)$ as the solution of the following boundary value problem:
\begin{equation}
	\label{eq:dynamic}
	\left\{\begin{array}{@{}l@{}}
		\nabla \cdot \mathbf{n}+l^{-2}_c\,\xi(r,t)-3\frac{Ca}{(R-r)} F(\delta)+\frac{H_a(r,t)}{\sigma}=0\\
		\partial_r \xi(R,t)=\mbox{ctg}(\theta (t))\\
		\partial_r \xi(0,t)=0\\
	\end{array}\right.\,
\end{equation} where $F(\delta)$ and $H_a(r,t)$ are correcting factors for the viscous and the inertial contribution. The first was proposed by \citet{delon2008relaxation} depending on $\delta=\mbox{ctg}(\partial_r h(r,t))$, with $\partial_r$ the derivative along the radial direction and $\mbox{ctg}()$ the cotangent. This term is \begin{equation}
F(\delta)=\frac{2}{3}\frac{\tan{\delta}\sin^2(\delta)}{\delta-\cos(\delta)\sin(\delta)}\,.\label{F_delta}
\end{equation}

The second was proposed by \citet{fiorini2022effect} and reads
\begin{equation}
   H_a(r,t)= \rho a_i(t) l_h \bigl(1-e^{-\frac{r-R}{l_i}}\bigr) \label{eq:pinertia}\,,
\end{equation} with $a_i (t)$ the instantaneous interface acceleration, $l_h= R c_t$ a characteristic length defined by the model parameter $c_t$, and $l_i (t)$ a model parameter controlling how the inertial forces decay towards the walls. 

The first correction accounts for the viscous dissipation near the contact line \cite{voinov1976hydrodynamics}, as the flow profile must comply with the no slip condition while still allowing for a moving contact line. The second correction accounts for the flow inertia, as the velocity profile far from the interface must adapt to the interface dynamics.

The solution to \eqref{eq:dynamic} provides the interface shape $\xi(r,t)$ from which the term $K(t)$ can be computed in \eqref{eq:Kterm} and inserted in \eqref{eq:utubedimensional}. 
At the limit $l_c/R\sim 1$, $Ca\rightarrow 0$ and $a_i\rightarrow 0$, equation \eqref{eq:dynamic} reduces to a spherical cap and the simplified theory with $K(t)=R\cos(\theta_D(t))$ is recovered. An alternative formulation, when the focus is placed on the modeling of the liquid column at the tube scale, is to introduce an equivalent contact angle $\theta_{D,m}(t)$ such that 
\begin{equation}
K(t)=R \cos \theta_{D,m}(t)\,.
\label{eq:macroscopicangle}
\end{equation}

The difference between the equivalent $\theta_{D,m}(t)$ and the actual $\theta_D(t)$ measures the discrepancy between the assumption of a spherical interface and the true interface.

In this work, the actual contact angles were fitted to a modified Voinov-Tanner law \cite{tanner1979spreading}, in which two unsteady terms are introduced to account for the history of the contact line and the contact-line acceleration $a_{cl}(t)$. Defining as $a^*_{cl}(t)=a_{cl}(t)/g$ the dimensionless acceleration, and $\theta = \theta_D^3-\theta_S^3$, with $\theta_S$ the static contact angle, the dynamic contact angle satisfies 
\begin{equation}
\theta(t)+\alpha\dot{\theta}(t) =\beta_1 Ca(t) + \beta_2 a^*_{cl}(t)
\label{eq:correlation}\, 
\end{equation} where $(\alpha,\beta_1,\beta_2)$ are empirical coefficients to be calibrated on the experimental data and the dot notation is used for time derivatives. Equation \eqref{eq:correlation} has the following analytical solution
\begin{equation}
\label{eq_sol}
\theta(t)=\frac{1}{\alpha}e^{- t/\alpha}\Bigl( \int^{t}_{0} (\beta_1 Ca(t')+\beta_2 a^*_{cl}(t') ) e^{ t'/\alpha}dt'  \Bigr)\,.
\end{equation}

This approach is a simplified version of the one proposed by \citet{bian2003liquid} and attempts to extend the traditional relationship between the contact angle and contact line kinematics to represent the data of this experiment. A much simpler model, which proved much more successful 
to describe the results of our experiments, is a simple linear trend of the form  
\begin{equation}
\theta_D(t) =\beta_3 Ca(t) +\theta_S 
\label{eq:correlationLinear}\, ,
\end{equation}

where the term $\beta_3$ can be seen has the inverse of Hocking contact line mobility coefficient \cite{ludwicki2022contact} which needs to be calibrated with the data. Similar linear relationship has been observed also by \citet{xia2018moving} for the case of sessile droplets on a oscillating support. The linear model is also known as Davis-Hocking correlation \cite{davis1980moving,hocking1987damping}.

\subsection{Scaling considerations}

The dimensionless form of equation \eqref{eq:utubedimensional} allows for analyzing the scaling laws of the different terms, and thus to position the results of this work in the literature of similar experiments. Moreover, the scaling analysis allows to evaluate the similarity with different fluids.

Denoting as $x^\ast=x/[x]$ the dimensionless scaling of a variable $x$ with respect to a reference $[x]$ and taking $[l]=l_c$, $[t]=(l_c/ g)^{1/2}$, $[u]=(l_c  g)^{1/2}$ and $[a]=g$ the reference length, time, velocity and acceleration, equation \eqref{eq:utubedimensional} becomes

\begin{equation}
\label{eq:utubenondimensional}
\begin{split}
     {\ddot{\bar{h}}}^\ast(t^\ast)= -8 C_f\frac{Oh_{cb}}{R^{\ast2}}{\dot{\bar{h}}}^\ast(t^\ast)
     -\frac{2}{L^\ast}{\bar{h}}^\ast(t^\ast)-\\
     \frac{2}{R^{\ast2} L^\ast}\left(K_A^\ast(t^\ast)-K_B^\ast(t^\ast)\right)\,,
    \end{split}
\end{equation} where $R^{\ast}=R/l_c$ and $L^{\ast}=L/l_c$ and $Oh_{cb} = \mu (g/\rho \sigma^3)^{1/4} $ is the Ohnsorge number based on the capillary length. This number solely depends on fluid properties and controls the viscous damping of interface oscillations. \citet{schmitt2015free} obtained a similar dimensionless equation using the radius of a cylindrical cell as characteristic length and a radius-dependent Ohnsorge number. 

Table \ref{tab:ohnsorge} collects the values of $Oh$, $R^\ast$, $L^\ast$ and the initial height $\overline{h}^*(0)$ for the experiments considered in this work together with experiments from \citet{fiorini2022effect} and \citet{dollet2020transition} as well as possible experiments using liquid hydrogen at $T=20$K, liquid oxygen at $T=90$K and liquid methane at $T=112$K, using the fluid properties reported in \citet{dreyer2007free}.

Interestingly, the similarity between liquid nitrogen and liquid oxygen is excellent. HFE7200 is in an acceptable similarity with nitrogen and oxygen but less with liquid hydrogen or methane. In case using water, the experiments of \citet{dollet2020transition} show comparable values of $R^*$ and $Oh_{cl}$ but not for $L^*$. However, for untreated glass surface, the stick slip contact line motion has been reported both by \citet{dollet2020transition} and \citet{fiorini2022effect} and the set of dimensionless numbers here presented might not provide a complete picture of the experiment. On the other hand, the experiment with ethanol \cite{dollet2020transition} has a similar dynamics to liquid nitrogen and HFE7200 \cite{fiorini2022effect}. However, the higher $Oh_{cl}$ and $R^*$ lead respectively to a higher damping due to viscous dissipation and a higher impact of the fluid inertia.

\begin{table}[h]
\centering
	\caption{\label{tab:ohnsorge} Comparison of the non-dimensional terms in equation \ref{eq:utubenondimensional} across different experiments. The properties of the cryogenic fluids are assumed as in \citet{dreyer2007free} with $R, L,$ and $ \bar{h}(0)$ values respectively 3.5, 100 and 20 mm as in this work.}
		\begin{tabular}{>{\centering\arraybackslash}p{3.7cm} c  c  c  c}
            \hline
               & $Oh_{cl} \cdot 10^{3}$ & $R^*$ & $L^*$ & $\bar{h}(0)^*$ \\
			\hline
		  This work (LN2 at $\approx 77K$)  & $1.9 $ & $3.3$ & $94.3$ & $18.9$\\
            HFE7200 in \cite{fiorini2022effect} & $4.9 $ & $4.1$ & $80$ & $10.2$\\
            Water  in \cite{fiorini2022effect} & $2.3 $ & $1.5$ & $29.4$ & $3.7$\\
            Water  in \cite{dollet2020transition} & $2.3 $ & $3.0$ & $39.1-53.9$ & $3.7$\\
            Ethanol in \cite{dollet2020transition} & $7 $ & $4.9$ & $86.9$ & $5.9$\\
            Liquid Hydrogen at $20K$  & $0.84$ & $2.1$ & $59.2$ & $11.8$\\
            Liquid Methane at $112K$ & $1.1 $ & $1.9$ & $54.6$ & $10.9$\\
            Liquid Oxygen $90K$ & $1.5 $ & $3.3$ & $93.5$ & $18.7$\\
		\end{tabular}
\end{table}

\renewcommand{\arraystretch}{1.1}
\begin{table}[h]
\centering
	\caption{\label{tab:properties} Fluids and U-tube physical properties.}
		\begin{tabular}{cc}
			\hline
			density $(\rho)$ & $812 \pm 2$ $\mbox{kg/m}^3$  \\
            dynamic viscosity $(\mu)$ & $0.176 \pm 0.002$$\mbox{mPa}\cdot \mbox{s}$  \\
            surface tension $(\sigma)$ & $9.23 \pm 0.15$$\mbox{mN/m}$  \\
            static contact angle $(\theta_S)$ & 0$\mbox{deg}$  \\
            tube radius $(R)$ & $3.5\mbox{mm}$  \\
            liquid column length $(L)$ & $99\pm2\mbox{mm}$  \\
		\end{tabular}
\end{table}


\section{Methodology}\label{sec:methodology}

\subsection{U-tube test case and Cryostat Facility}

\begin{figure}[h]
\centering
\includegraphics[width=8.5cm,clip]{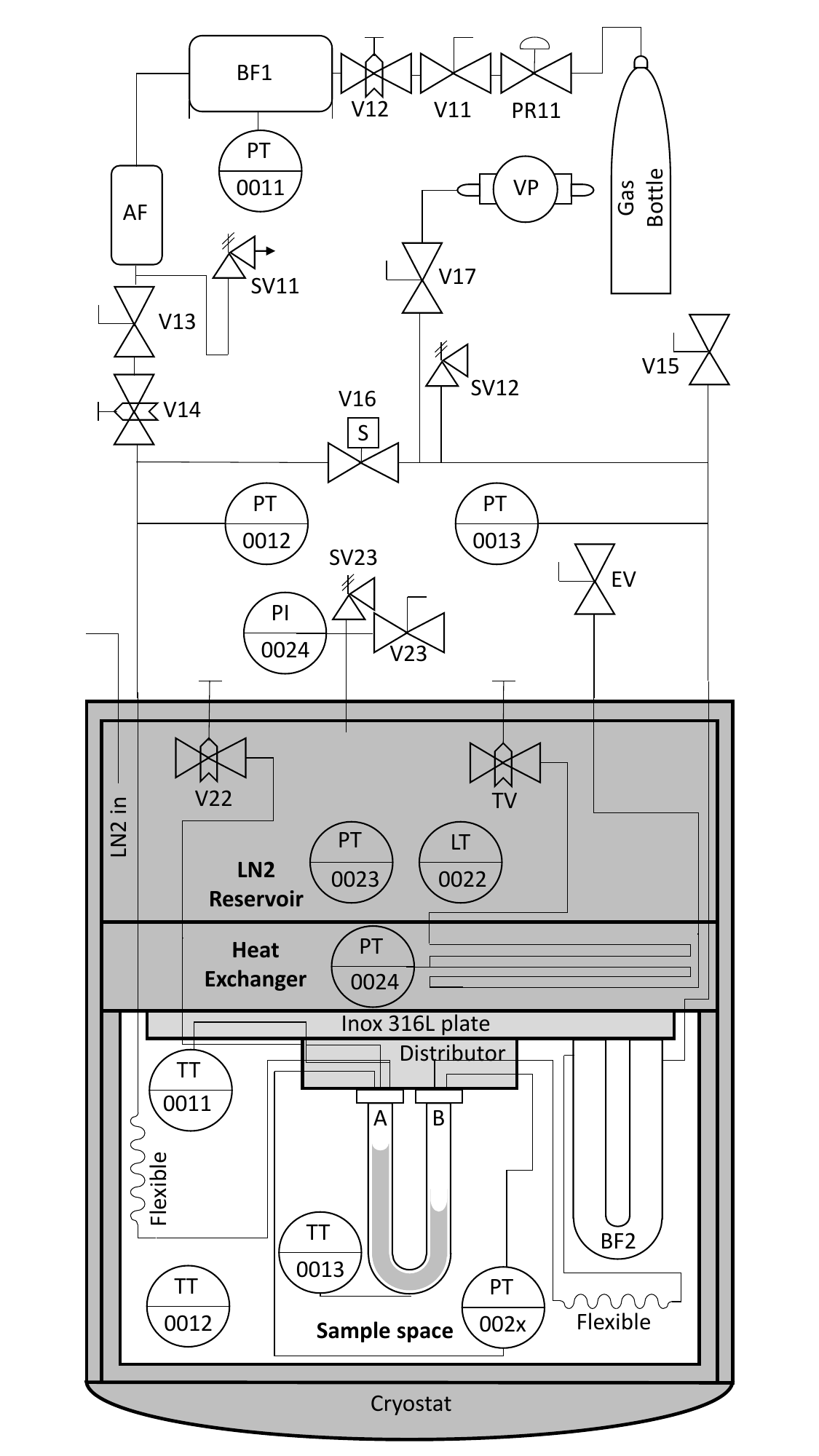}
\caption{Schematic of cryostat facility. Pressure sensors are indicated with PT and Pressure Indicators with PI. Temperature sensors are indicated with TT and level transducers with LT. }
\label{fig:PandID}       
\end{figure}

The experiments were carried out at the cryostat facility from the von Karman Institute. A schematic of the facility and the connections between its components is shown in Figure \ref{fig:PandID}. The U-tube sides are labelled A and B. The line connected to the side A is controlled via valves V14, V13 and air filter (AF) and the buffer tank (BF1) where the pressure is set using the pressure regulator PR11 and the valves V11 and V22. This is the gas feeding line, which controls the initial pressurization in the experiment, and is connected to gas bottles. Side B is connected to the gas-discharge line via valve V15 or via valve V17. The first discharges in atmosphere while the second is connected to a vacuum pump VP.

 A fast response cryogenic ball valve (Triad series 60C, V16 in Figure \ref{fig:PandID}) controls the connection between the two sides, allowing for separating the ullage gas on the two sides once the curved side of the tube is filled with the test liquid. Safety valves SV11 and SV12 are placed on each line. The cryostat consists of annular volumes with the U-tube at the center in the sample space (white area in Figure \ref{fig:PandID}). The reservoir of the cryostat is filled with liquid nitrogen, which flows through a serpentine heat exchanger and undergoes phase change by throttling (through valve TV). 
 The vaporization of liquid nitrogen cools the heat exchanger contact gas block which then cools the cryostat’s sample space. The nitrogen vapor is vented to the ambient atmosphere depending on the cooling rate needed. The throttle valve TV and a nitrogen exhaust valve EV allow controlling the vapor pressure and thus the cooling power of the cryostat. The pressure in the heat exchanger is monitored with the Kulite CTL-190 pressure transducer PT-0024. The cryostat is cooled constantly as long as the nitrogen is replenished. The steady-state is obtained by matching the cooling power from the nitrogen with the heat loss.

\begin{figure}[h]
\centering
\includegraphics[width=7.5cm,clip]{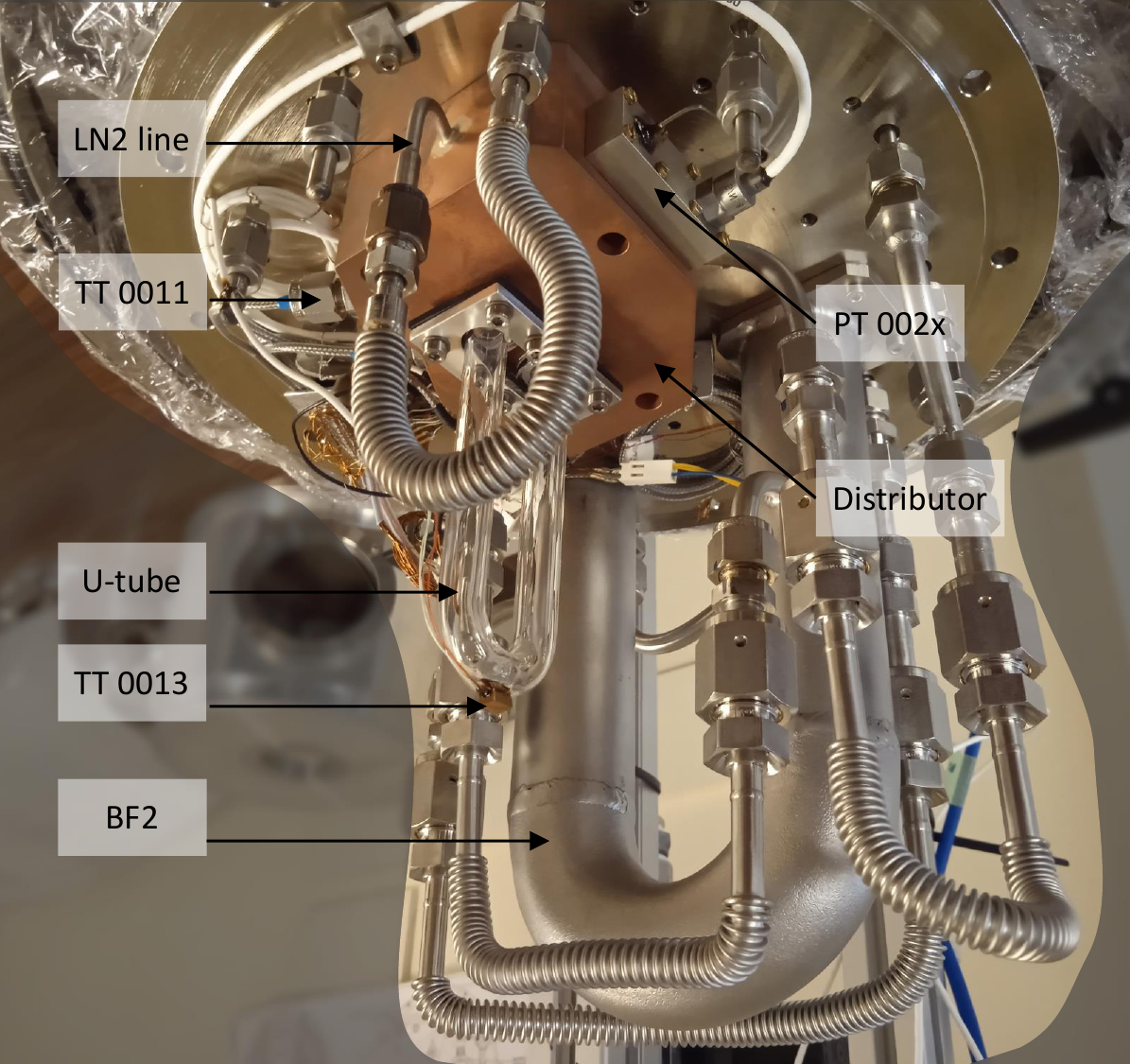}
\caption{Picture of the U-tube and the channel distributor assembly, showing the U-tube together with some of the components shown in the schematic of Figure \ref{fig:PandID}.}
\label{fig:Distributor}       
\end{figure}

Figure \ref{fig:Distributor} shows a picture of the channel distributor that connects the U-tube to the cryostat. The distributor connects to the cryostat through Inox 316L plate, 8 Inox 316L 3.5$\varnothing$ mm screws and spring-energized O-rings (Fluolion 01 Virgin PTFE). Figure \ref{fig:PandID} shows both the distributor and the U-tube positions in the cryostat’s sample space with the Inox 316L plate in contact with the heat exchanger. The distributor has two $ \sfrac{1}{4}$'' male Swagelok VCR fitting brazed to its side to connect with an external gas input and output line. A third $\sfrac{1}{8}$'' VCR allows connecting with an internal channel that transfers the liquid nitrogen from the nitrogen reservoir through the distributor and inside the U-tube. We use the buffer volume BF1 outside the cryostat to achieve fine control of the pressure on the input gas-line. A second buffer volume, labeled as BF2, is positioned inside the cryostat connected to side B of the U-tube and allows for adjustment of the pressure response of the system to approximate a step response. 

The entire setup and all fittings were helium leak checked before testing. Before starting the filling procedure, all the volumes are flooded with helium vapor and later pumped down to a pressure of $10^{-3}$ Pa to remove any traces of ambient air and water vapor inside the cryostat. The purging cycle is repeated three times, and a small amount of helium is added to act as a heat exchange gas between the sample space and the heat exchanger.  

The motion of the liquid column is produced by setting an initial level difference between the two sides, pressurizing line A, then opening valve V16 to produce a step pressure reduction. The liquid oscillation lasts several seconds before the equilibrium position is recovered. The initial pressurization was achieved using pure Nitrogen gas bottles filling tube A with V16 closed. 

The input gas line starts from the gas cylinder visible in Figure \ref{fig:PandID}. The pressure is set to a few hundred Pascals in the outer buffer tank BF1, where pressure is monitored via Pressure Transducer PT-0011. The air filter AF removes impurities in the buffer, and the gas is released in the test cell via valve V13, while V14 is used to regulate flow. The input gas line cools down as it flows through the entire cryostat liquid nitrogen reservoir before reaching the sample space. We added flexibles to each side of the U-tube to increase the heat exchange of the input gas with the sample room and achieve similar conditions with the gas already present in the channel distributor. During the experiments, we monitored the gas temperature on side A, and we observed a temperature variation $\Delta T<0.25K$ upon the pressurization.

Figure \ref{fig:PandID} shows the three temperature sensors connected to the test cell. There are: (1) a Lakeshore silicon diode DT-670 (TT-0011) mounted on the copper distributor of the test cell by use of SWAGELOCK connectors, (2) a second Lakeshore DT-670 (TT-0013) in contact with the external bottom part of the tube via cryogenic glue and (3) a Lakeshore RTD Cernox CX-1050-AA-HT-1.4L (TT-0012) suspended in the helium exchange gas close to the test cell. The temperature sensors are connected to a Lakeshore Model 218 temperature controller for data logging and processing. The corresponding calibration curves for temperature sensors are built into the temperature controller. The pressure in the test cell is monitored by three pressure sensors. The pressure in the test cell is monitored using three pressure sensors, also shown in figure \ref{fig:PandID}. These are two miniature ruggedized pressure sensors Kulite CTL-190 (PT-002x and PT-0012) and an AMS 5812-0150-D pressure sensor (PT-0013). The sensor PT-002x is connected to side A with the reference port connected to side B (differential configuration) while sensor PT-0012 is connected to the input gas line. Both PT-002x and PT-0012 signals are conditioned through MICRO ANALOG 2 – FE-MM4 module. The sensor PT-0013 is connected to the gas venting port close to the discharge valve. This sensor was calibrated using a standard Druck pressure calibrator and used to adjust the manufacturer calibration for PT-0012 and PT-002x at the temperatures encountered in our experiment. The recalibration of these sensors was performed with the sensors mounted at the respective locations and increasing the pressure in the U-tube using the gas-feeding line A.  In the case of PT-002x, the calibration requires that the test cell is partially filled with the test liquid. The calibration offset is acquired with the gas-liquid interfaces resting at the same height, while the sensor sensitivity is acquired by slowly increasing the pressure on side A of the tube with V16 closed. The output voltage is acquired when stable conditions of the interface are achieved and compared with the readings from PT-0012 and PT-0013.

\subsection{Experimental procedure}

At equilibrium starting conditions, the liquid interfaces on the two sides of the U-tube have the same height. The gas volumes of sides A and B are isolated by closing the cryogenic valve V16. The initial conditions are imposed by moving the interfaces out-of-equilibrium pressurizing side A, setting a level difference between the two interfaces, and ensuring their position is stable (zero initial interface velocity). The initial over-pressure level is controlled in the external buffer tank by the pressure regulators V14 and V13 (see Figure \ref{fig:PandID}).
The interfaces remain stable at the selected height for a few seconds, during which opening valve V16 triggers the experiment. If V16 remains closed, condensation of the introduced gas nitrogen begins on side A, slowly reducing the overpressure and moving the two interfaces back to equilibrium. After the pressurization phase (max 300 Pa), the temperature of the ullage gas corresponds to the value in saturated conditions at the operating pressure of the tube. In this work, we performed experiments with the gas temperature in the range $74.2-74.8 K$, where differences between experiments are due to the high sensitivity of the cryostat to the environmental conditions, the amount of liquid nitrogen remaining in its reservoir and the number of experiments performed. 

After opening valve V16, the liquid column oscillates freely around the equilibrium position. For each experiment, we record the motion of one of the two interfaces to maximize the interface resolution. We use a high-speed camera (model JAY SP-12000-CXP4), acquiring grey-scale images at 300 fps. The interface shape is obtained by casting an image of the shadow of the meniscus on the camera using diffused light source on the opposite side of the tube. Two optical access of 75$\varnothing$ mm give access to the sample space of the cryostat facility. The active region of the camera is restricted to the central region of size 4096x768 pixels to achieve the highest acquisition frequency allowed by the camera. The camera mounts a 105mm lens and it is positioned to acquire the motion of the interface spanning the full tube length.

\subsection{Interface tracking and contact angle measurement}

The interface tracking was carried out via image processing combining edge detection with correction for optical distortion and regression of the interface model in \eqref{eq:dynamic} on the detected interface position. The methodology is extensively described in \citet{fiorini2022effect}, to which the reader is referred for more details. The regression consists in identifying the model parameters $l_h, l_i$  and $\theta$ in \eqref{eq:dynamic} in order to allow for robust computation of the contact angle. The regression was solved using the Nelder-Mead algorithm \cite{nelder1965simplex}, implemented in the Python library scipy.optimize \cite{2020SciPy-NMeth}. The same tool is used for the regression of the dynamic contact angle correlation \eqref{eq:correlation} with the experimental dynamic contact angle data.

The fitting requires the accurate determination of the liquid–wall interface location. The images should be well aligned with the tube's wall since a small misalignment can result in significant errors in the contact angle measurement \cite{bellur2016contact}. The image pixel size was measured by taking the tube's external diameter as a reference is $15\pm0.5\mu$m as measured by averaging from 10 images and 5 randomly chosen locations.

The image-based interface detection provides the first points at a distance of about $15-30\mu$m from the wall. The regression is carried out using the available points, and $\bar{h}(t)$ is obtained using Equation \ref{eq:h_average}. An important aspect, however, is that the boundary value problem in \eqref{eq:dynamic} requires the contact-line velocity (through the Capillary number) as an input to compute the interface profile. This computation could be carried out iteratively, starting from the mean interface velocity as a guess and then adjusting from the interface solution. However, such an approach is particularly sensitive at the extremely low contact angle considered in this work because small variations of $\theta_D$ produce large variations of the contact-line position $h(R,t)=h_{CL}(t)$.

In the literature \cite{iliev2011dynamic,maleki2007landau,petrov1993quasi}, these difficulties have been circumvented by imposing the Capillary number from the average relative velocity between interface and wall. A similar approach has been followed by \citet{dollet2020transition} in the same U-tube configuration considered in this work: these authors take the average interface velocity as the contact-line velocity. In recent works on oscillating droplets \cite{xia2018moving,brunet2022unstationary}, the contact-line position was identified from image analysis, and the contact-line velocity computed via time differentiation, while authors working on static sessile droplets with HFE7100\cite{kumar2020sessile,garivalis2022experimental} circumvented the problem by engraving a pinning location on the substrate.

\begin{figure}[ht]
    \centering
    \includegraphics[width=8.5cm,clip]{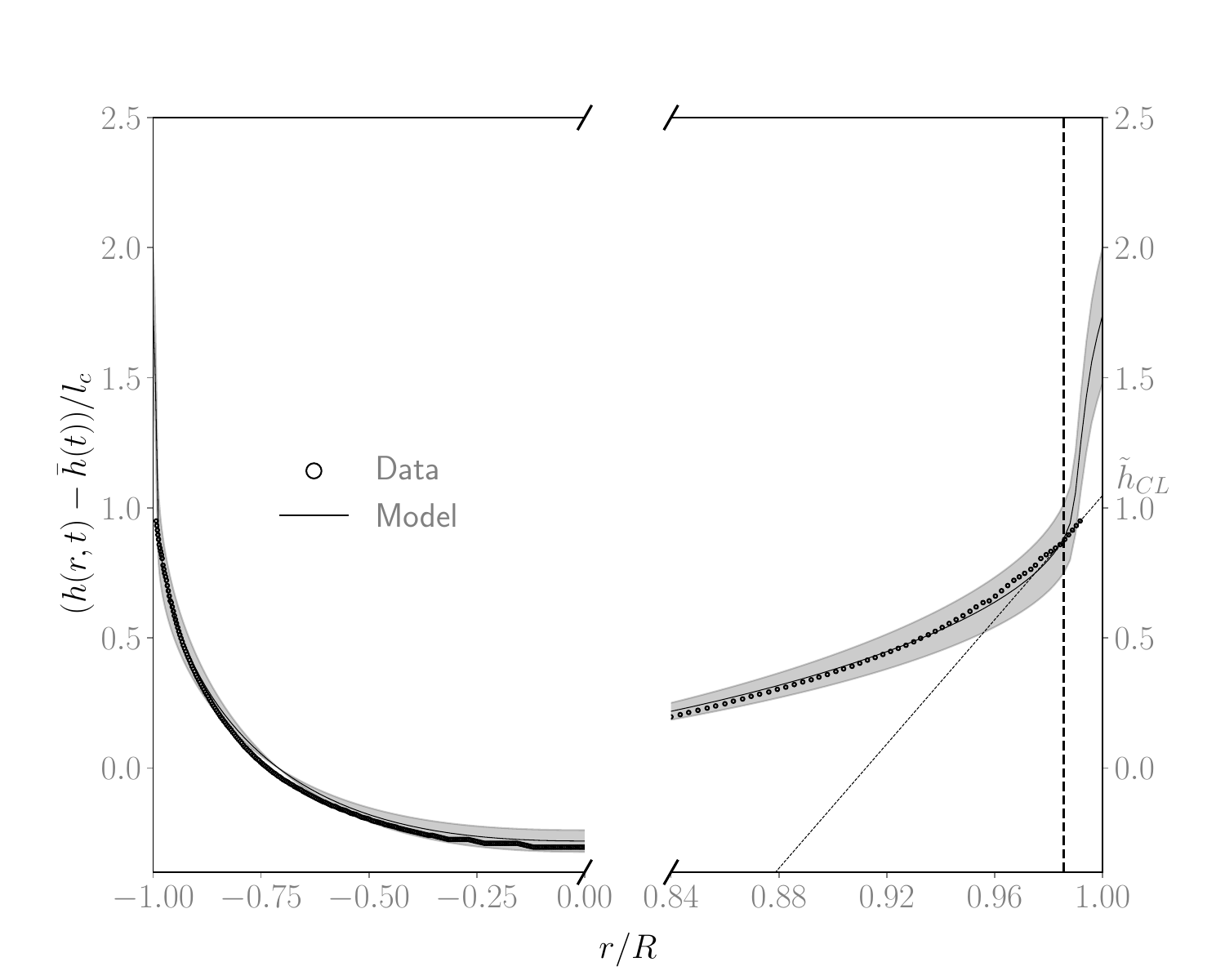}
    \caption{The portion on the left shows the results of interface regression on the data. The portion on the right shows a zoom near  the wall, plotting the interface regression together with data and the linear extrapolation from which the contact-line velocity is defined. The model-based measurement yields $\theta_D = 4.16 \pm 0.51^o$. }
    \label{fig:regression}
\end{figure}

Our work combines the challenge of a time-varying contact-line velocity with near zero contact angle, and we seek to separate the oscillations of the average interface motion from the local motion of the contact line. We thus compute the Capillary number required in equation\eqref{eq:dynamic} from an approximation of the interface at a `larger scale' than the interface model provides. This is obtained by linearly extrapolating the interface prediction from a distance of $50\mu$m from the wall. An example of interface tracking, model regression, and local extrapolation is shown in Figure \ref{fig:regression}, along with the uncertainty calculation carried out via the Monte Carlo approach presented in \citet{fiorini2022effect}. The portion on the left shows a `large scale' view of the interface regression, which closely matches the image-based detection of the interface. The portion on the right shows a zoom near the wall, along with the linear extrapolation carried out for $r<R-50\mu$m and the contact line position $\tilde{h}_{CL}$ at its intersection with the wall. The interface model generally predicts a much higher interface location, i.e. $h_{CL}>\tilde{h}_{CL}$, thus a smaller contact angle. However, the computation of the contact line velocity by time differentiation of $h_{CL}(t)$ is too sensitive to the small ($\pm 0.5^o$) uncertainties in the contact angle computation.

Therefore, in what follows, we compute the contact-line velocity (and thus the Capillary number required in \eqref{eq:dynamic}) by time differentiation of $\tilde{h}_{CL}$. This is used to solve the regression of the interface model in \eqref{eq:dynamic} and the resulting interface was used to compute the dynamic contact angle $\theta_D$.

 
\section{Results and Discussions}\label{sec:results}

We present and discuss the contact angle and interface measurements in section \ref{sec:res1}. Section \ref{sec:res2} focuses on the relation between contact angle and interface dynamics by introducing an equivalent macroscopic contact angle. Finally, section \ref{sec:res3} closes with a note on the relative importance of the forces driving the investigated configuration.

\subsection{Contact Angle and Interface Dynamics}\label{sec:res1}

We consider two types of experiments to characterize the interface dynamics on both sides of the tube using only one camera. These are denoted as `advancing' and `receding' experiments and differ in the initial condition. In the `advancing', the interface starts with $\overline{h}(0)=-18.4\pm1.5$mm. Thus the interface evolves along a dry surface during the first rise (when it is in advancing conditions) while it evolves over a pre-wet surface afterward. On the contrary, in the `receding' experiments, the interface starts with $\overline{h}(0)=18.4\pm1.5$mm. Thus the interface recedes in the first descent and advances on a pre-wet surface afterward. Both types of experiments start with a still interface and with the release of the overpressure on side A of the tube (see section \ref{sec:methodology}). The two experiments are dynamically identical and should provide the same column oscillation up to a sign change. The videos in each experiment are processed to provide the history of the average interface height $\bar{h}(t)$, the evolution of the interface shape $h(r,t)$ and the contact angle evolution $\theta_D(t)$. All experiments are repeated three times to assess repeatability.

\newcommand{\mysize}{0.18}

\begin{figure*}[ht!] 
	\begin{minipage}[l]{0.49\linewidth}
		`Advancing' Experiment\\
		\vspace{1ex}
		\centering
		\captionsetup[subfigure]{labelformat=empty}
		\begin{subfigure}[t]{\mysize\textwidth}
			\includegraphics[width=\textwidth]{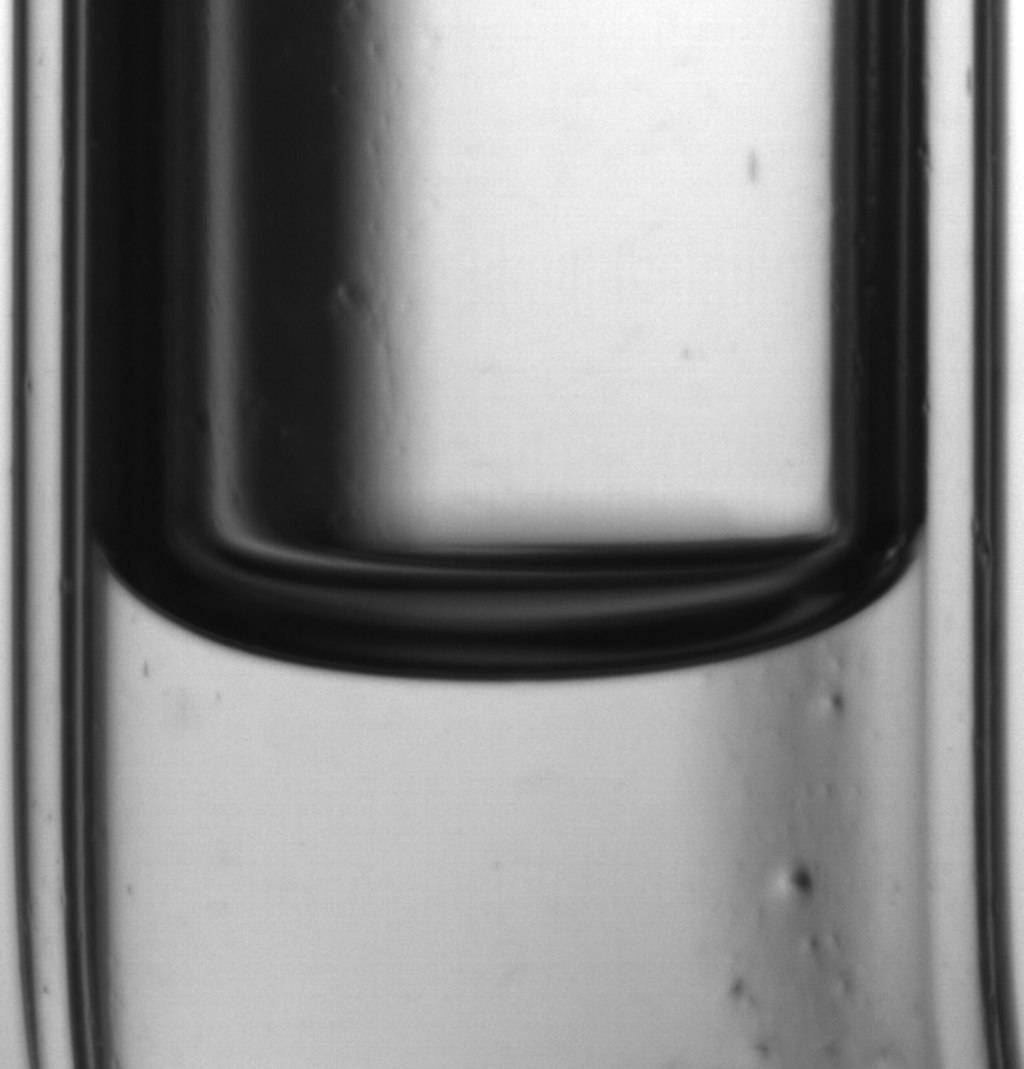}
			\caption{t=0s}
		\end{subfigure}
		\begin{subfigure}[t]{\mysize\textwidth}
			\includegraphics[width=\textwidth]{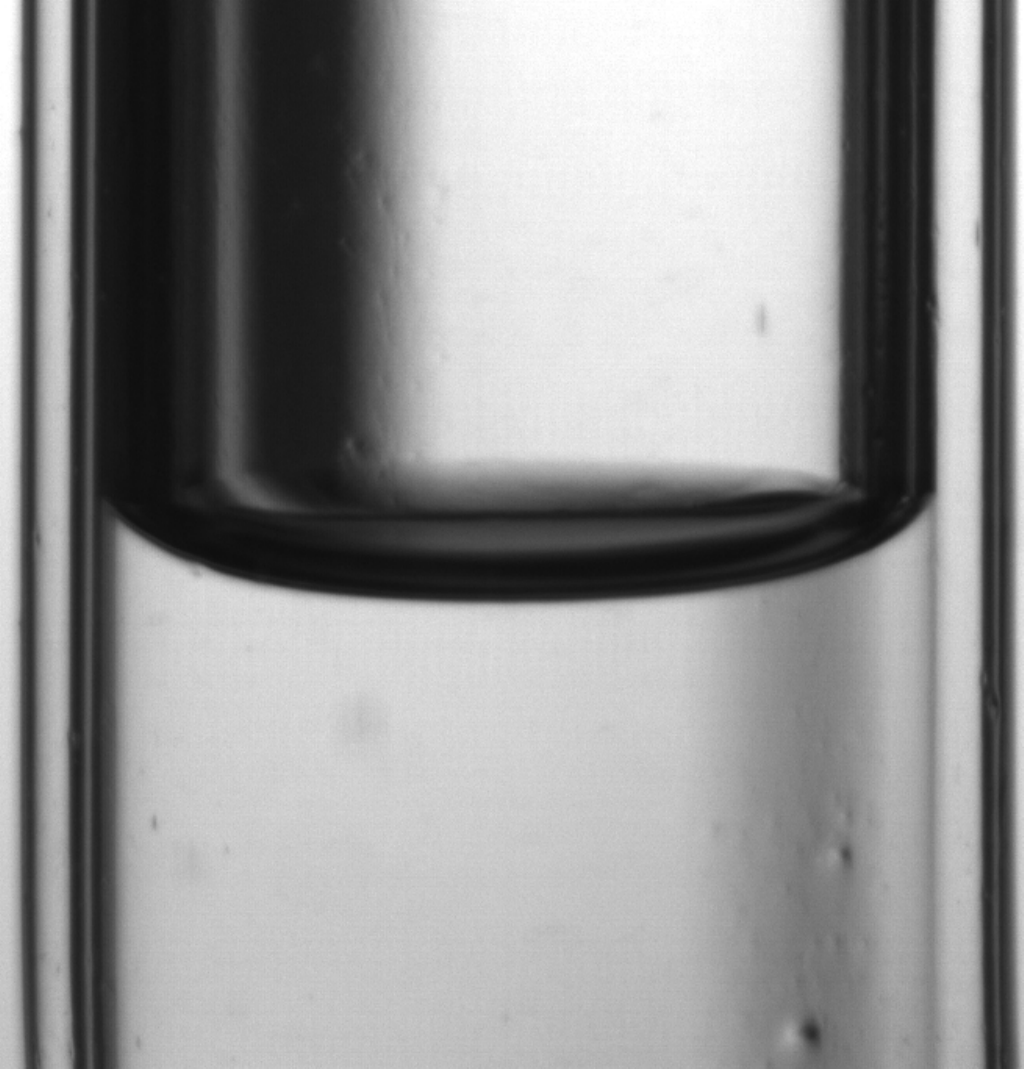}
			\caption{t=0.07s}
		\end{subfigure}
		\begin{subfigure}[t]{\mysize\textwidth}
			\includegraphics[width=\textwidth]{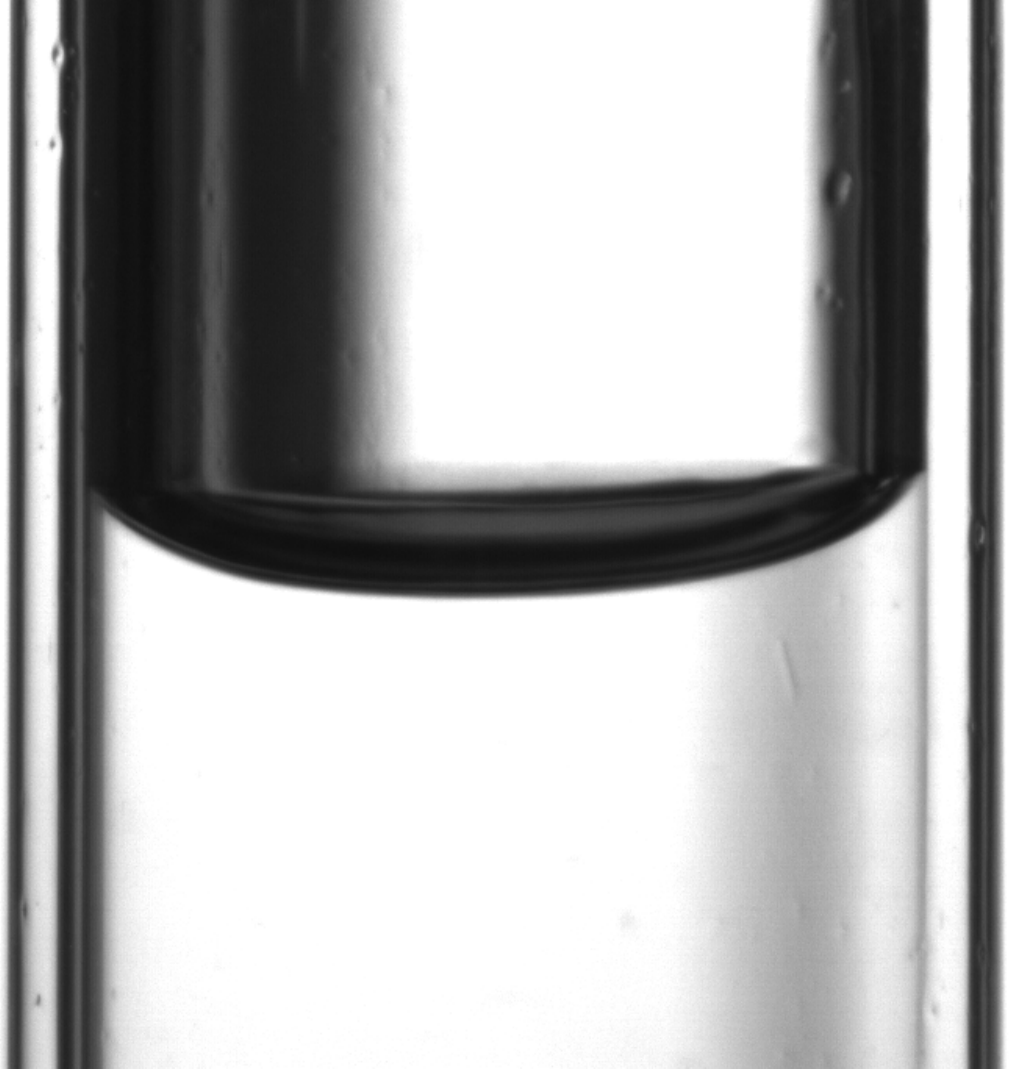}
			\caption{t=0.13s}
		\end{subfigure}
		\begin{subfigure}[t]{\mysize\textwidth}
			\includegraphics[width=\textwidth]{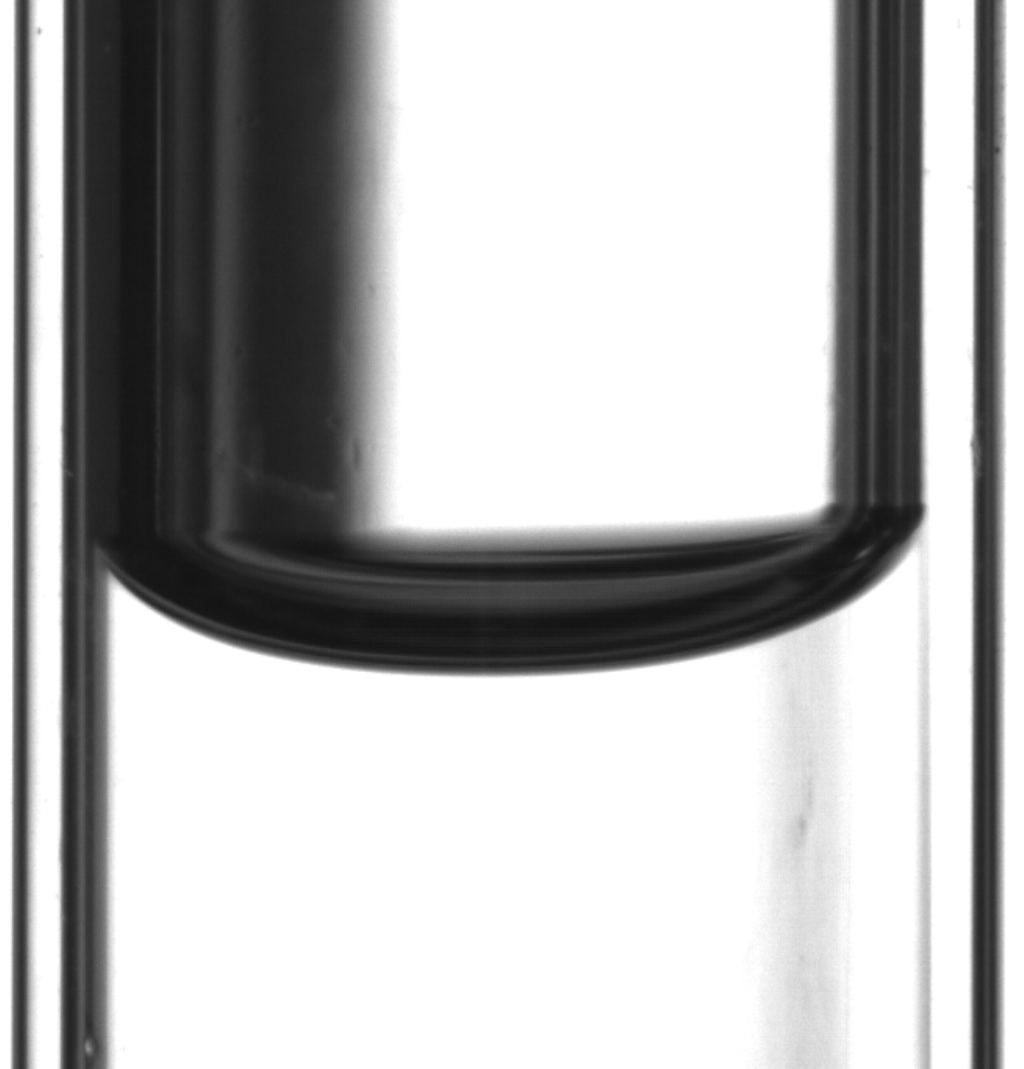}
			\caption{t=0.2s}
		\end{subfigure}
		\begin{subfigure}[t]{\mysize\textwidth}
			\includegraphics[width=\textwidth]{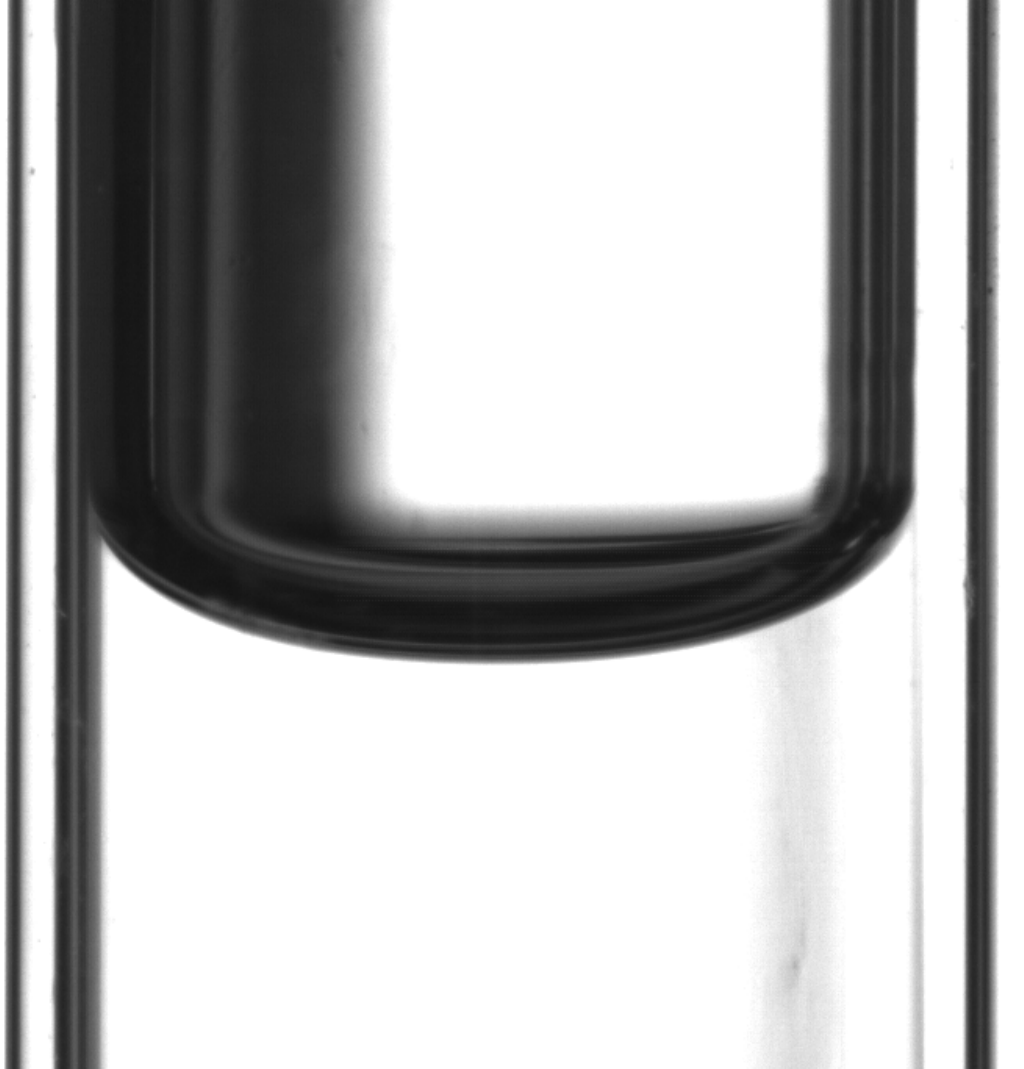}
			\caption{t=0.27s}
		\end{subfigure}\\
		(a)\\
	\end{minipage}
	\vline
	\begin{minipage}[r]{0.49\linewidth}
		'Receding' Experiment\\
		\vspace{1ex}
		\centering
		\begin{subfigure}[t]{\mysize\textwidth}
			\includegraphics[width=\textwidth]{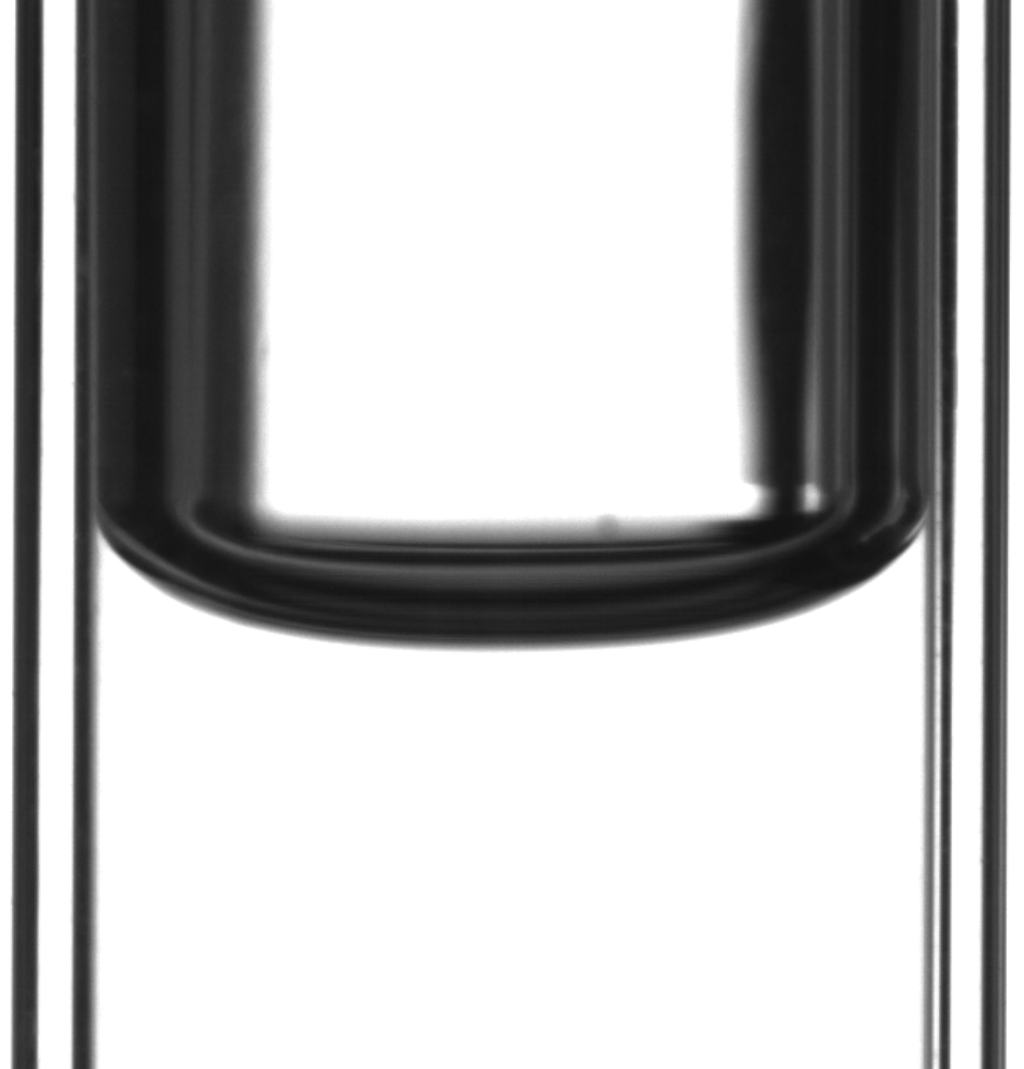}
			\caption*{t=0s}
		\end{subfigure}
		\begin{subfigure}[t]{\mysize\textwidth}
			\includegraphics[width=\textwidth]{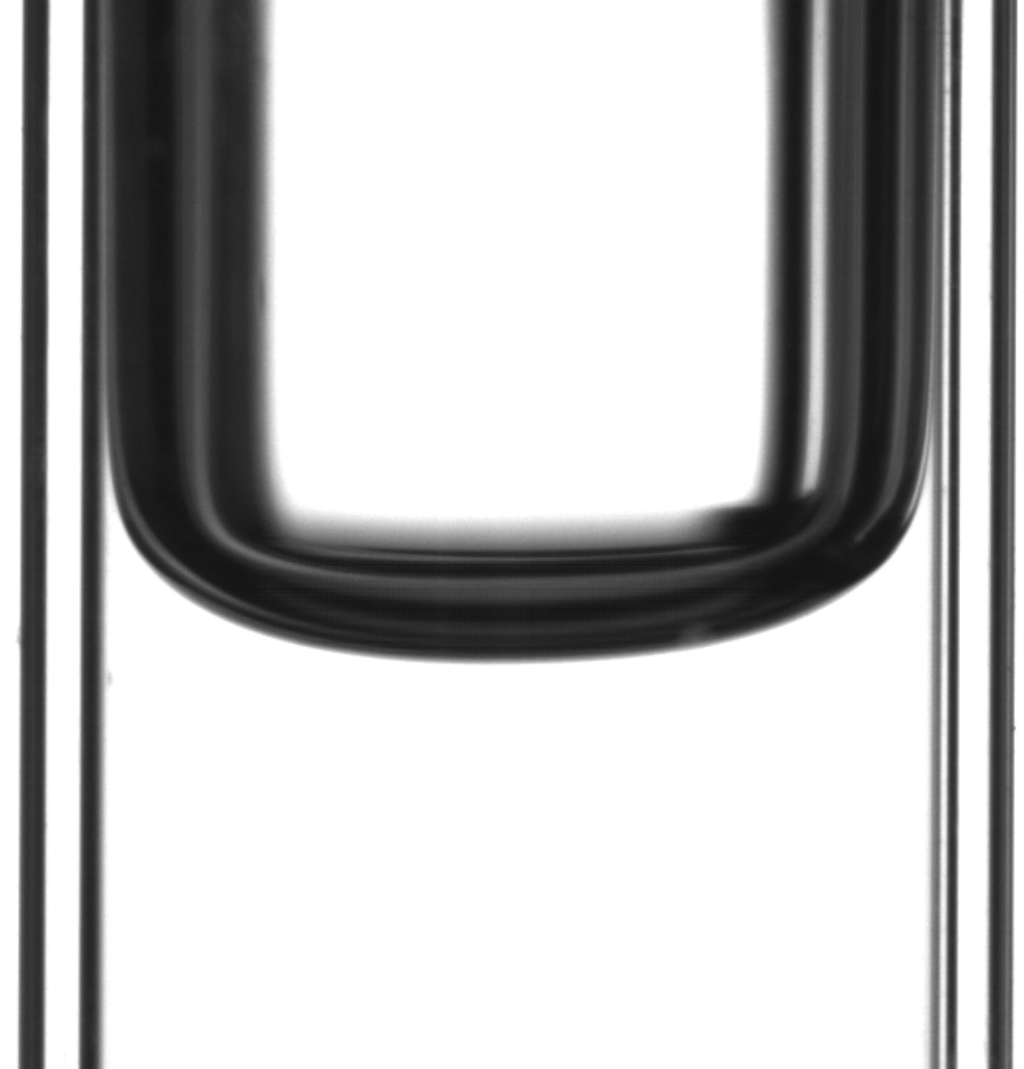}
			\caption*{t=0.07s}
		\end{subfigure}
		\begin{subfigure}[t]{\mysize\textwidth}
			\includegraphics[width=\textwidth]{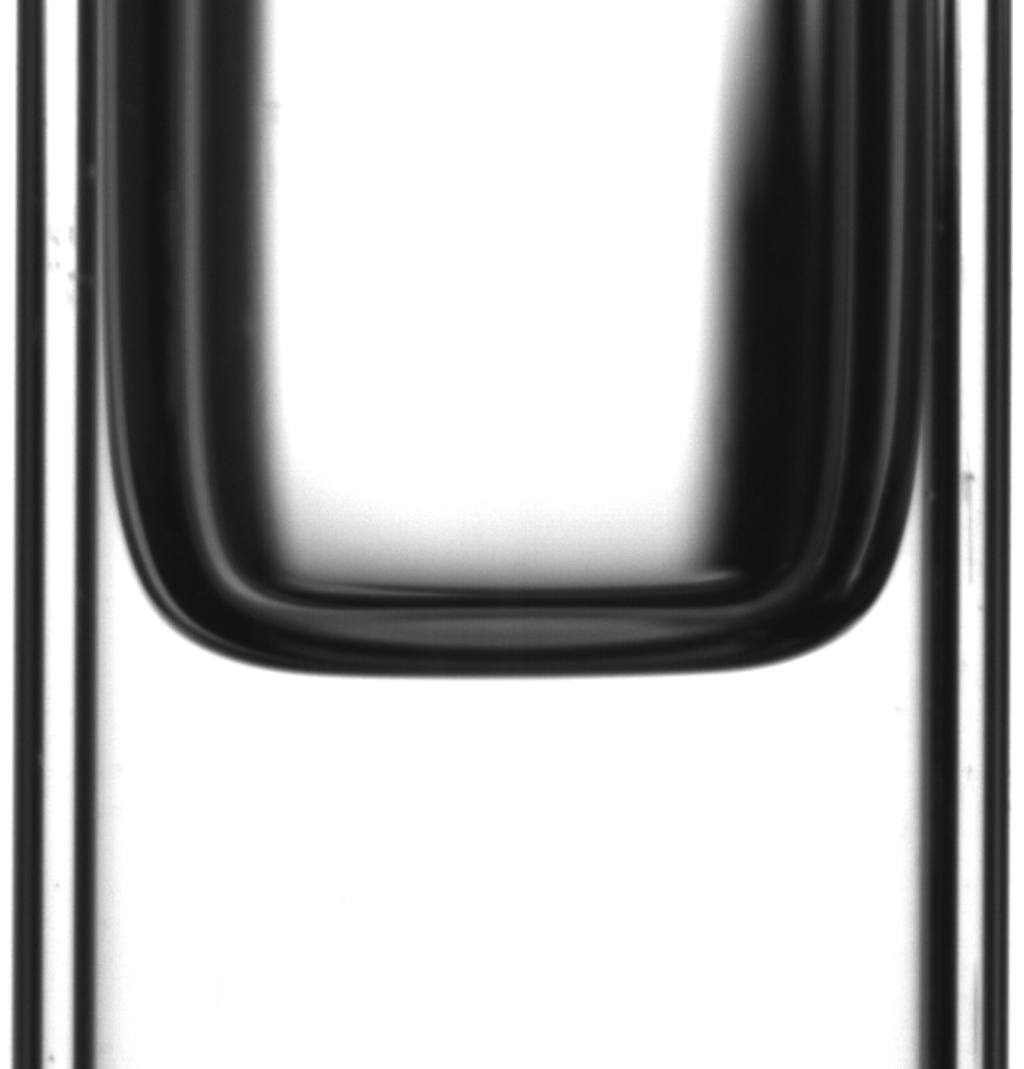}
			\caption*{t=0.13s}
		\end{subfigure}
		\begin{subfigure}[t]{\mysize\textwidth}
			\includegraphics[width=\textwidth]{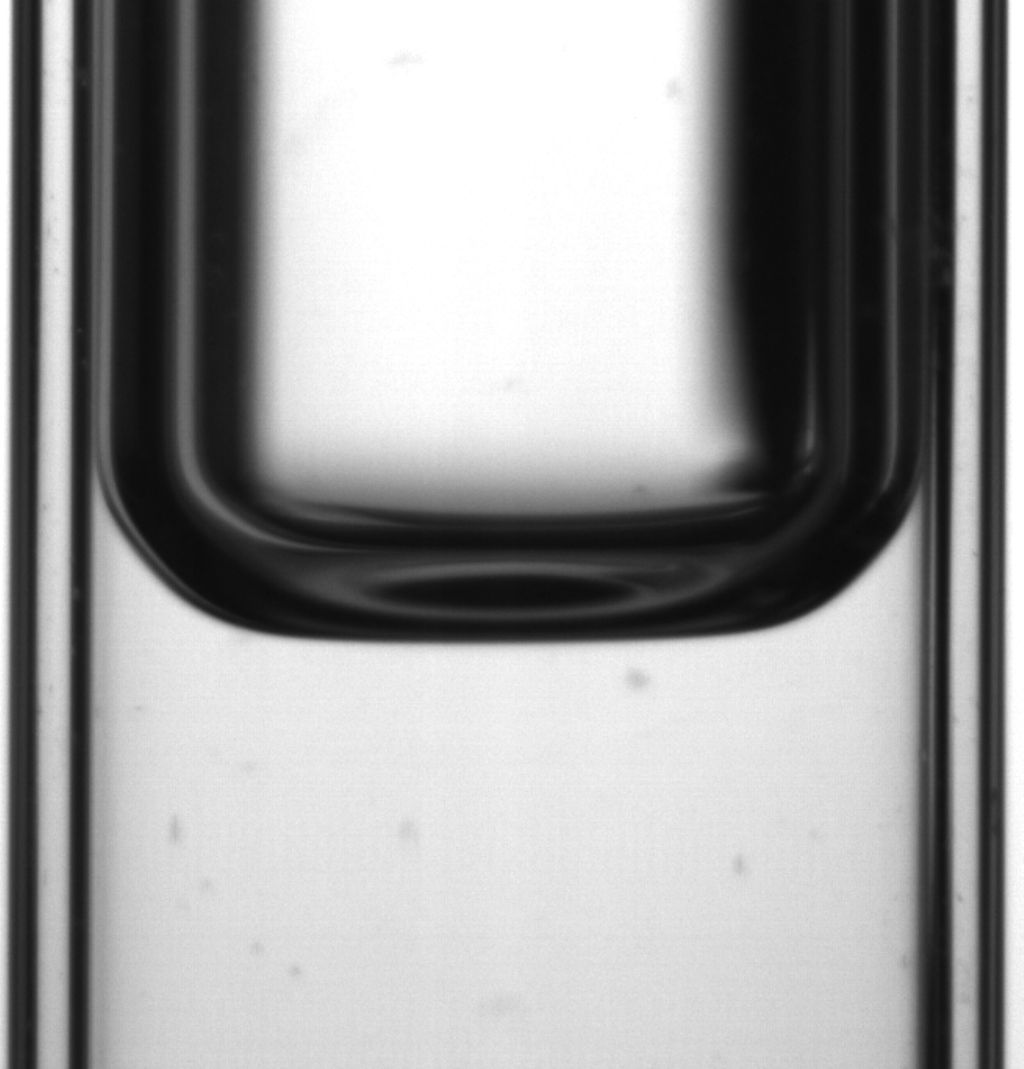}
			\caption*{t=0.2s}
		\end{subfigure}
		\begin{subfigure}[t]{\mysize\textwidth}
			\includegraphics[width=\textwidth]{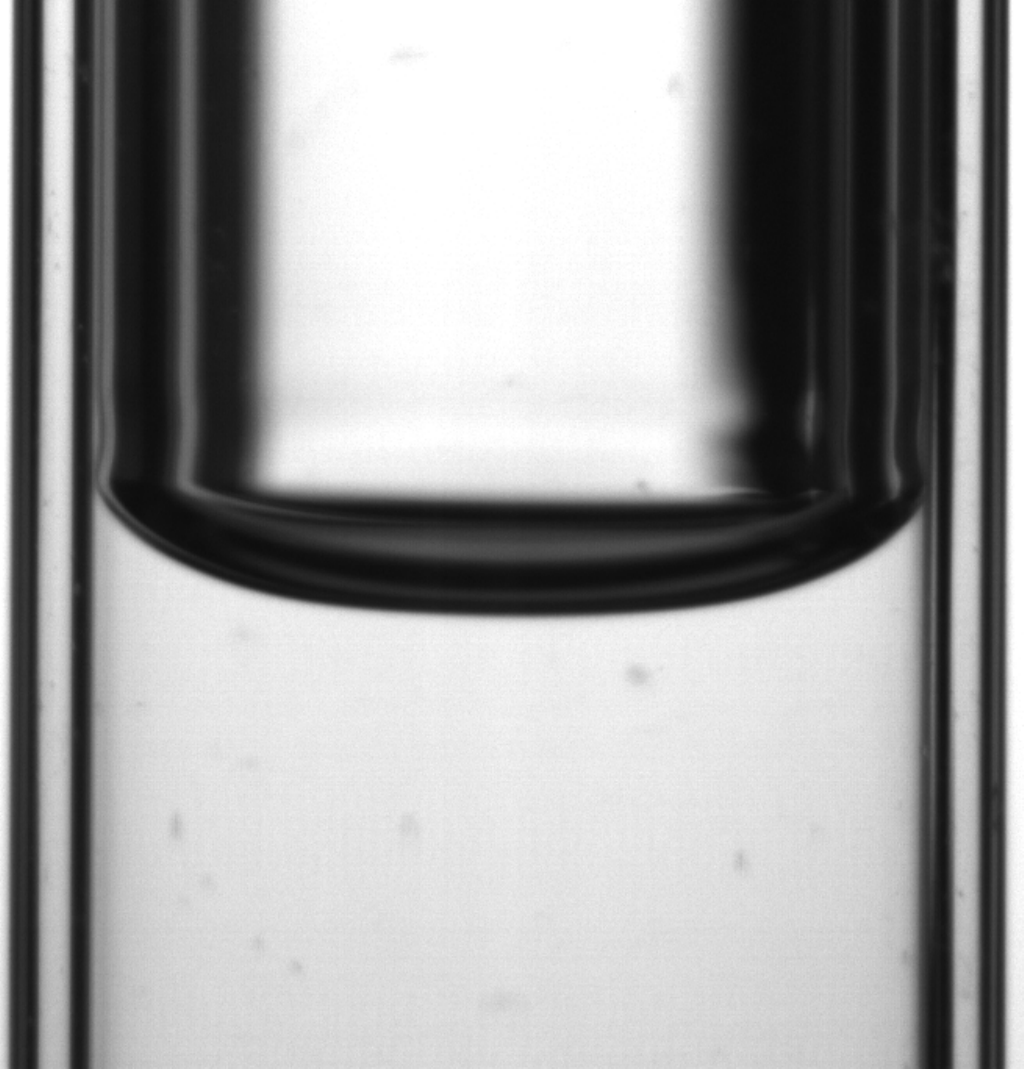}
			\caption*{t=0.27s}
		\end{subfigure}\\
		(b)\\
	\end{minipage}
	\begin{minipage}[l]{0.49\linewidth}
		\vspace{1ex}
		\centering
		\includegraphics[width=8cm,clip]{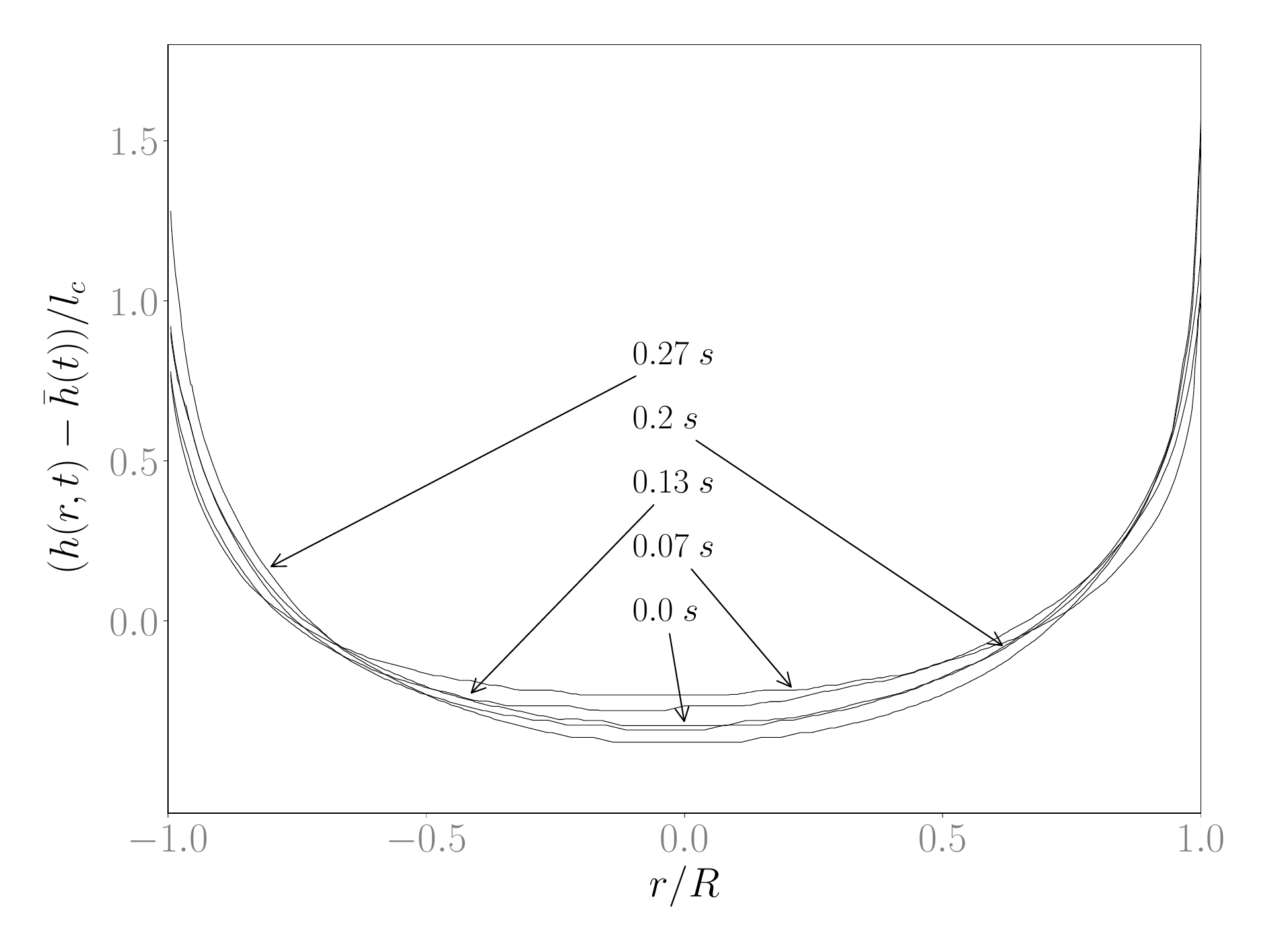}\\
		(c)\\
	\end{minipage}
	\vline
	\begin{minipage}[r]{0.49\linewidth}
		\vspace{1ex}
		\includegraphics[width=8cm,clip]{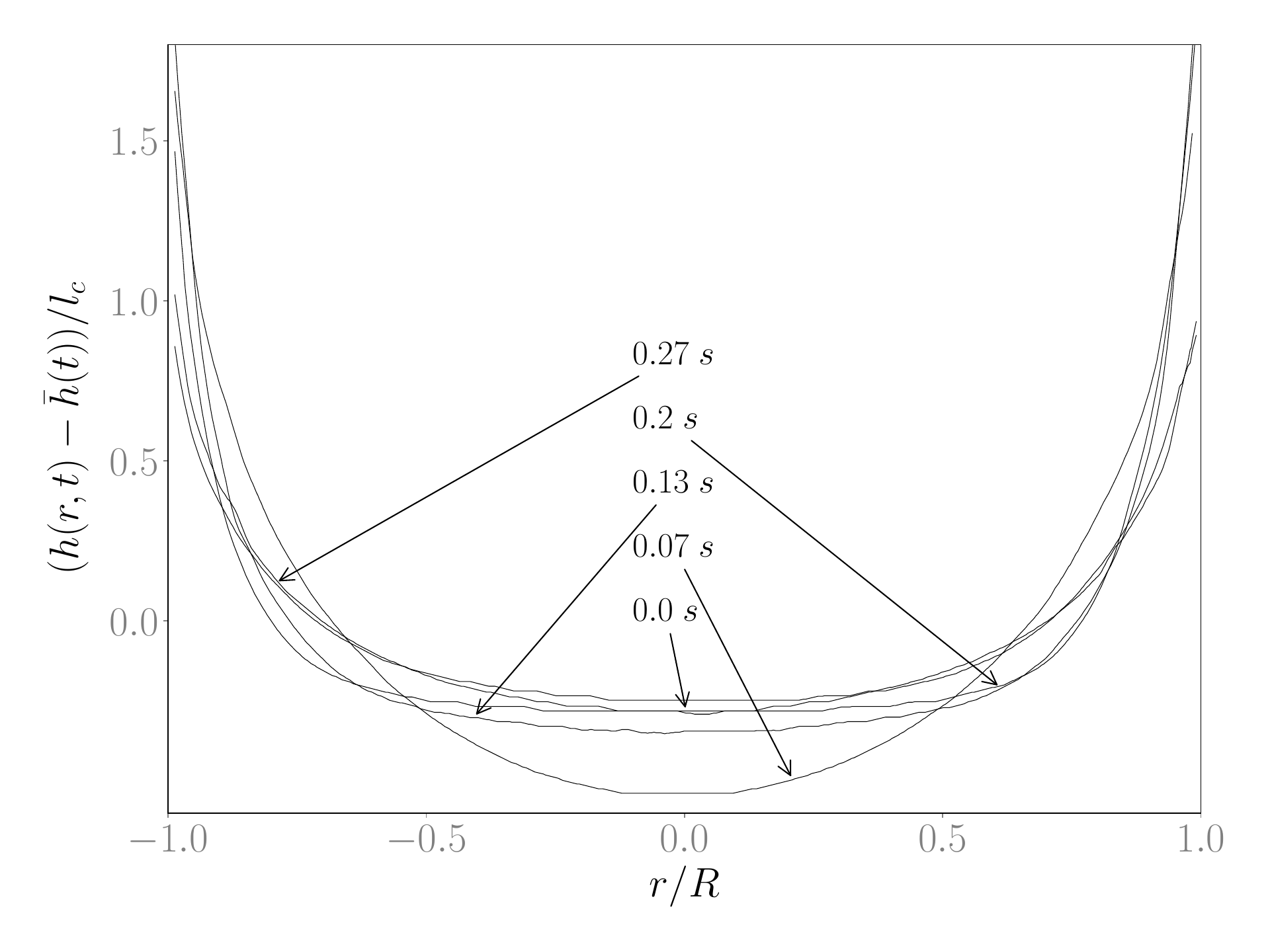}\\
		(d)\\
	\end{minipage}
	\begin{minipage}[l]{0.49\linewidth}
		\vspace{1ex}
		\centering
		\includegraphics[width=8.8cm,clip]{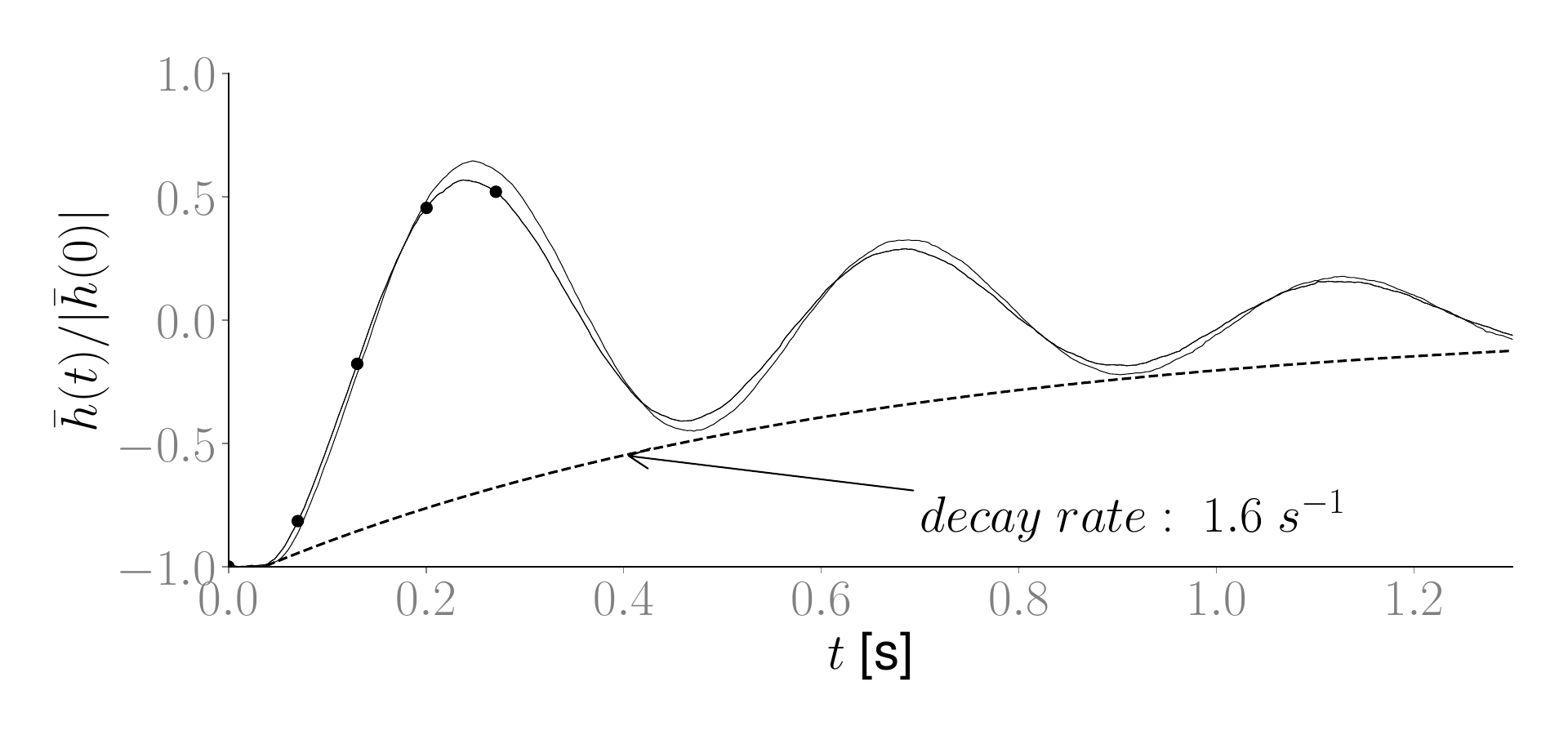}\\
		(e)\\
	\end{minipage}
	\vline
	\begin{minipage}[r]{0.49\linewidth}
		\vspace{1ex}
		\includegraphics[width=8.8cm,clip]{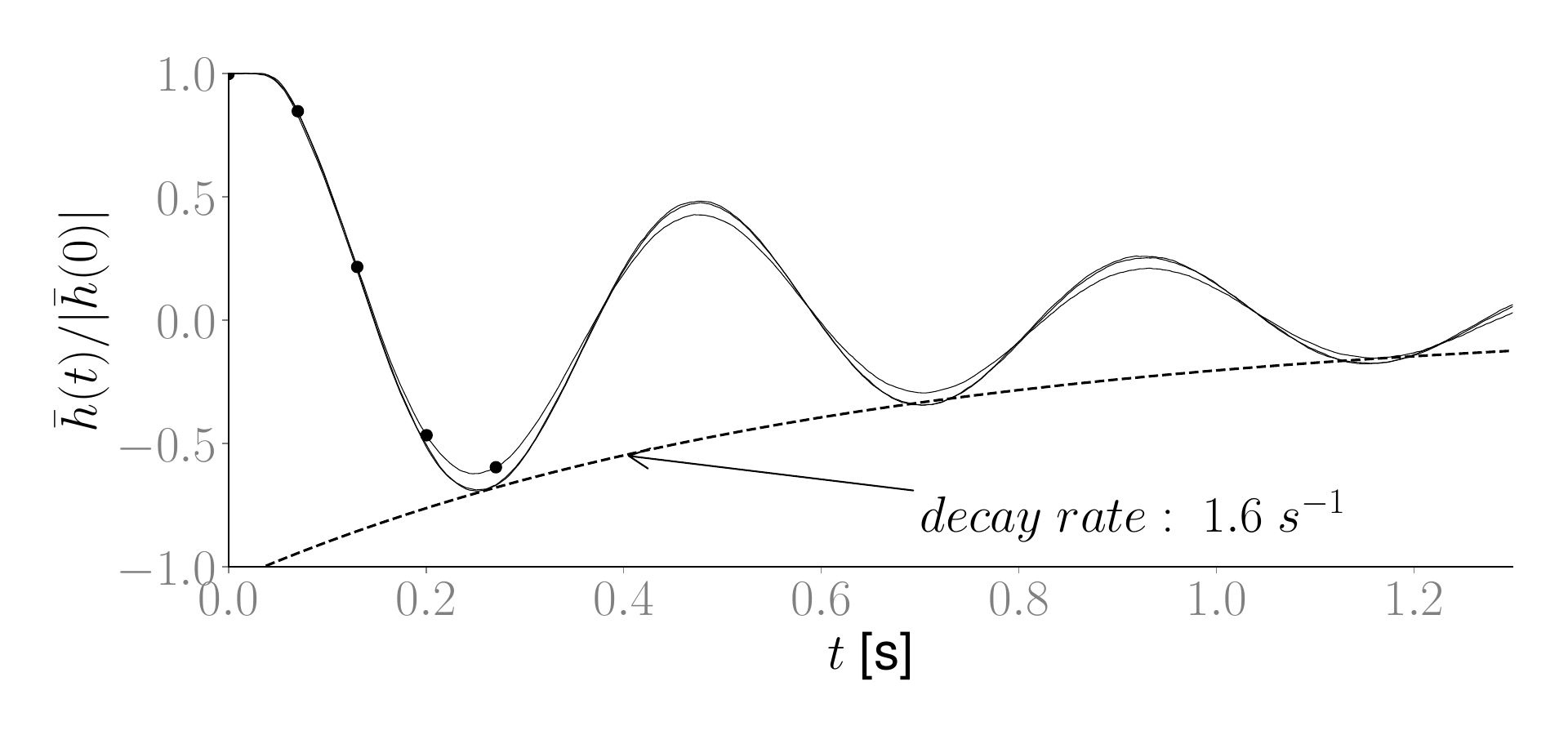}\\
		(f)\\
	\end{minipage}
	\begin{minipage}[l]{0.49\linewidth}
		\vspace{1ex}
		\centering
		\includegraphics[width=9cm,clip]{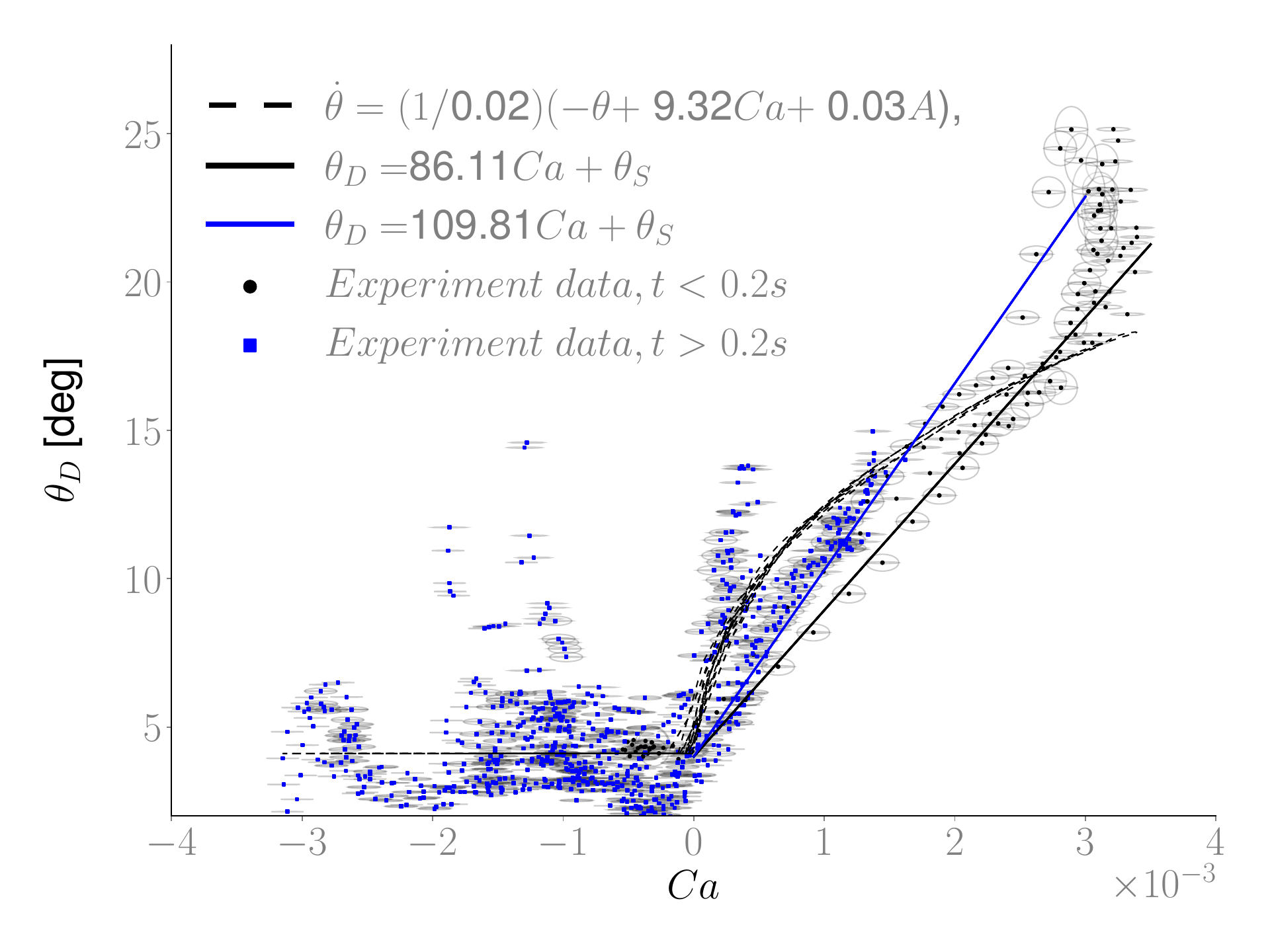}\caption*{(g)\label{fig:experimentadvancing}}
	\end{minipage}
    \hspace{0.1ex}
	\vline
	\begin{minipage}[r]{0.49\linewidth}
		\vspace{1ex}
		\includegraphics[width=9cm,clip]{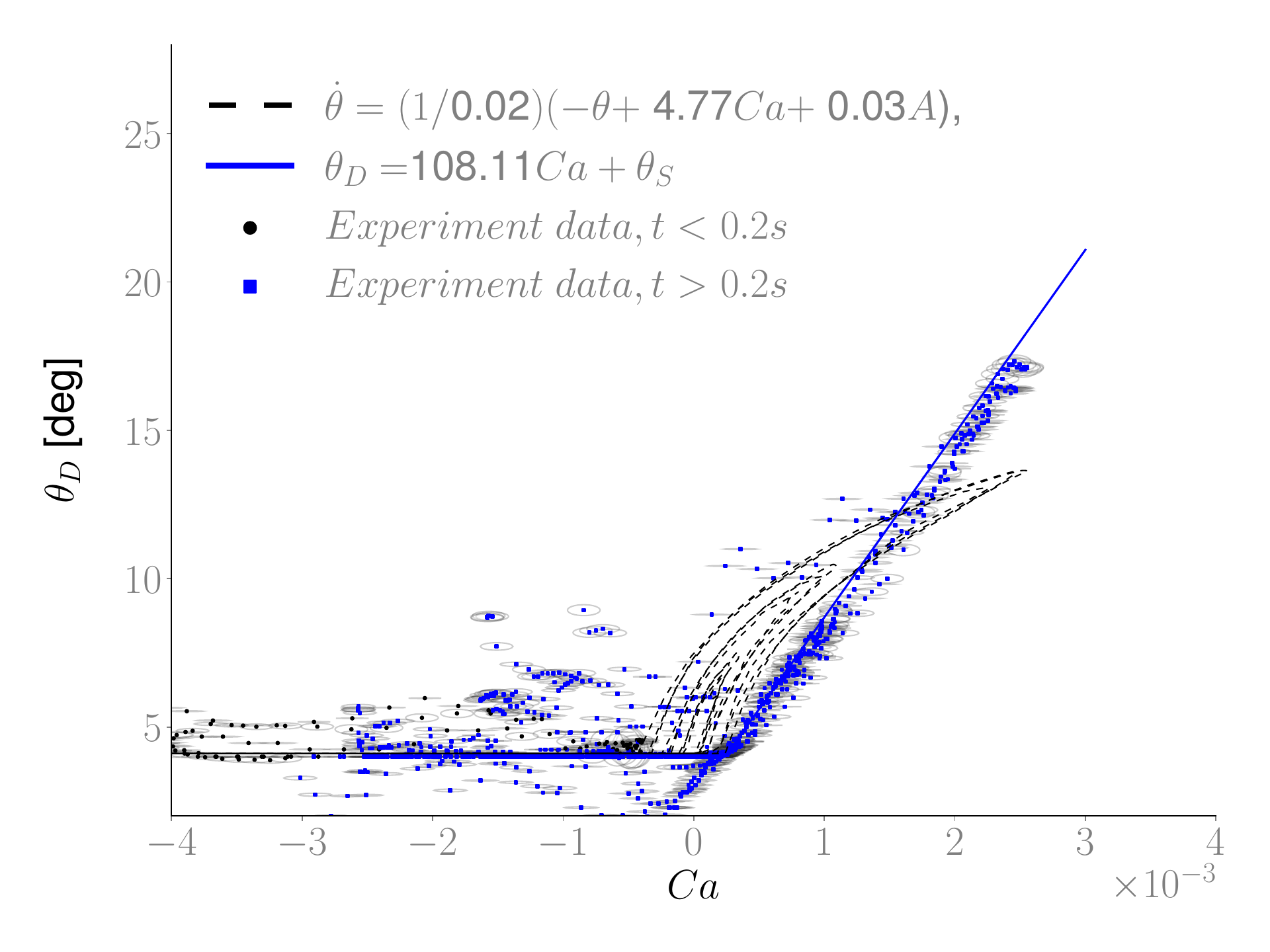}\caption*{(h)\label{fig:experimentreceding}}
	\end{minipage}
	\vspace{3mm}
	\caption{Summary of the measurements in advancing (figures on the left) versus receding (figures on the right) conditions. Figures (a) and (b) shows some selected snapshots. Figures (c) and (d) shows the corresponding interface detection while figure (e) and (f) shows the time history of average interface. Figures (g)-(h) plot the dynamic contact angle versus Capillary number. }  	
\end{figure*}

We summarize the main results for both tests in Figure 5. The figures on the left column refers to the advancing experiment while the figures on the right refers to the receding experiment. Figures 5(a) and 5(b) show five snapshots of the video recording for each set of conditions, cropped near the interface. From a visual inspection and from a close zoom in the images, no film is visible in the receding phase. Yet, as we shall see, the dynamics of the contact angle is different when evolving on a dry or a pre-wet surface. Figures 5(c)-5(d) show the interface detection for the same snapshots. For plotting purposes, the curves are shifted to have zero spatial average. In all tests the interface remains symmetric during the entire experiment but is far from a spherical.

Figures 5(e)-5(f) show the history of the average interface $\bar{h}(t)$ for the three runs carried out for each kind of experiment. Only two lines are distinguishable because two of the tests lead to identical curves; these are nevertheless kept because the contact angle measurements in all runs were considered for the regression of the contact angle correlations. In these plots, the dashed line shows the exponential envelope of the oscillation maxima and minima. This turns out to be $\propto e^{-\lambda t}$, with the decay rate $\lambda=1.6$ s. The excellent agreement between the 'advancing' and 'receding' experiments is further proof of the measurement repeatability. Moreover, the exponential envelope of the oscillation shows that the dynamics of the liquid column in the U-tube, as described by \eqref{eq:utubedimensional}, is well approximated by a linear second-order system. This was also the case of the experiments presented by \citet{fiorini2022effect} for HFE7200 and water. The last set of plots (g and h) shows the contact angle evolution against the Capillary number. In both plots, the dashed line shows the best prediction for model \eqref{eq:correlation}, with the coefficients $\alpha, \beta_1$, and $\beta_2$ identified via optimization on the three sets of test cases. Different markers are used to distinguish plots with $t<0.2$s (evolving on a dry surface) from those with $t>0.2$s (evolving on a wet surface). The circle around each point provides the 95\% confidence interval to account for the measurement uncertainty. This was computed fitting a bidimensional Gaussian distribution on the velocity and contact angle measurement. The continuous lines in both plots are used for the linear model in \eqref{eq:correlationLinear}. 

In both test cases, the modified Voinov-Tanner law in \eqref{eq:correlation} fails to reproduce the measured dynamics while the linear Davis-Hocking correlation \eqref{eq:correlationLinear} succeeds, at least in advancing conditions ($Ca>0$). In these conditions, the `advancing' experiments highlights the different contact angle evolution in the case of dry or pre-wet surfaces. In both cases, the linear relation holds, but the slope is different. The linear trend in the pre-wet conditions for both series of experiments is in good agreement. In receding conditions ($Ca<0$), regardless of the dry or pre-wet surface status, the dynamic contact angle remains much closer to the static one but less predictable.

\begin{figure*}[htb!]
\captionsetup[subfigure]{skip=-0.2\textwidth,margin=-0.3\textwidth}
\begin{subfigure}[t]{0.49\textwidth}
\centering
\includegraphics[width=7.5cm,trim =0 0 6cm 0, clip]{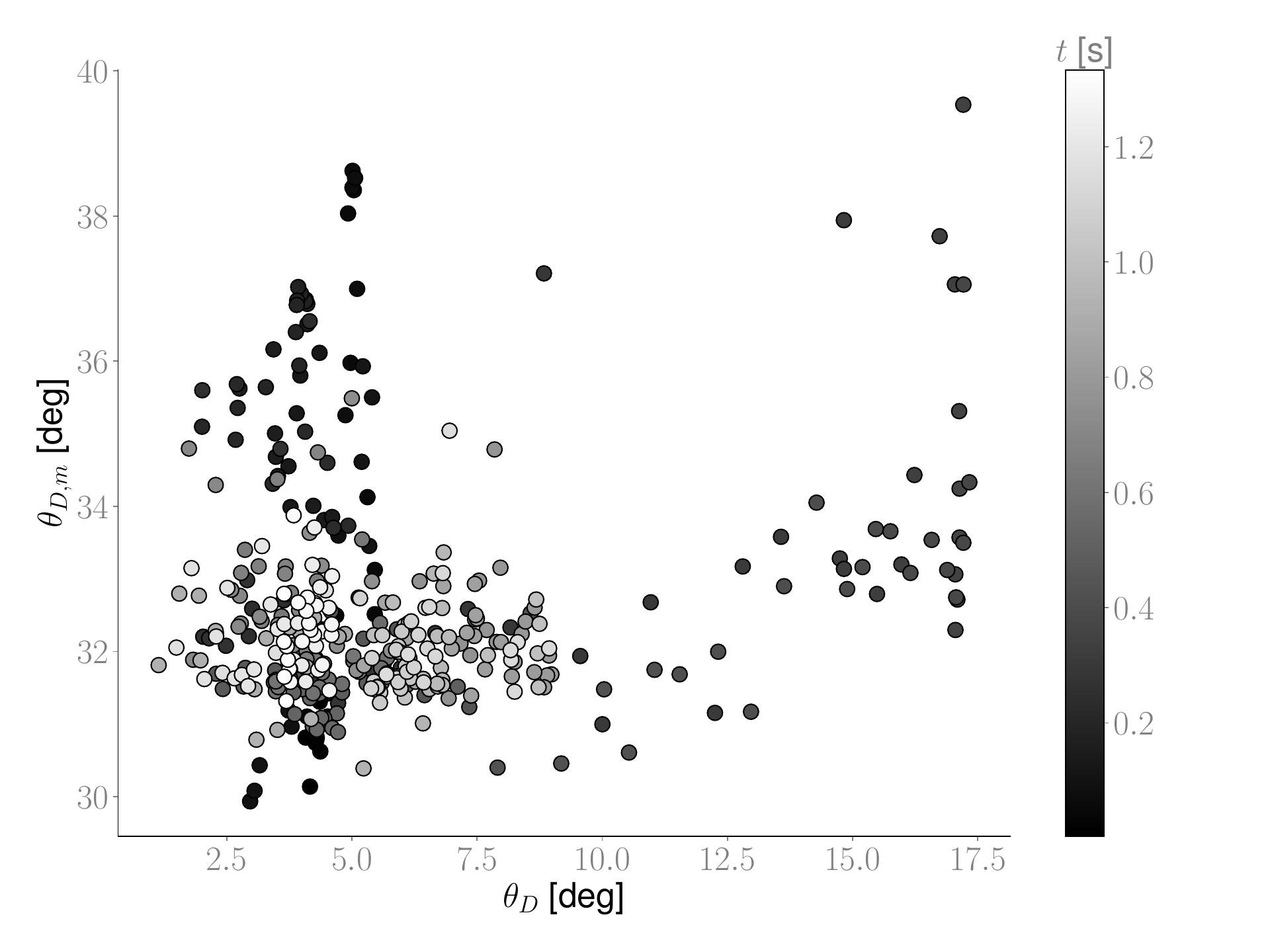}
\caption{\label{subfig:Krec}}
\end{subfigure}
\begin{subfigure}[t]{0.49\textwidth}
\centering
\includegraphics[width=9cm,clip]{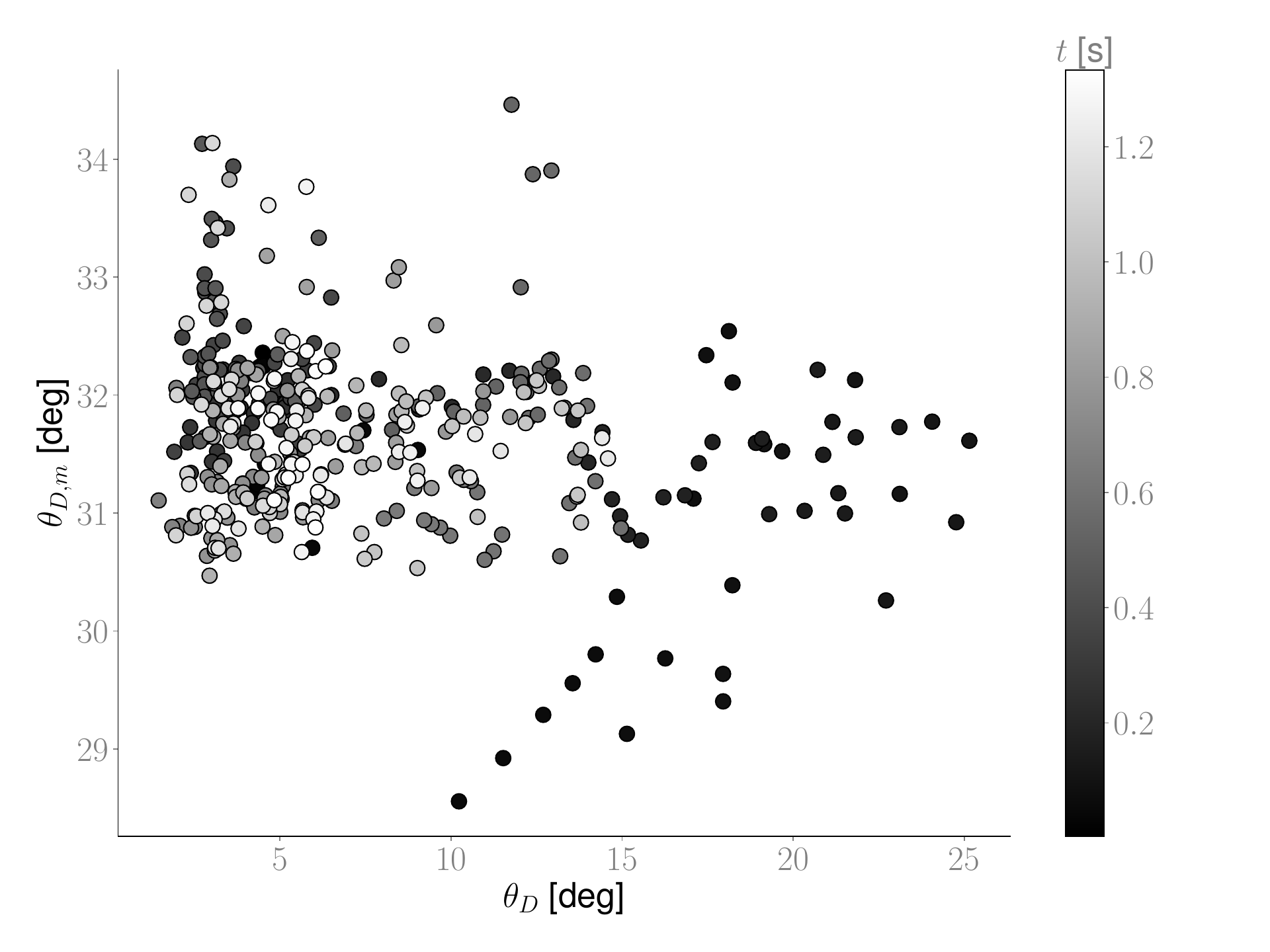}
\caption{\label{subfig:Kadv}}
\end{subfigure}
\caption{Comparison of the equivalent contact angle computed with Equation \ref{eq:macroscopicangle} and the measured dynamic contact angle. On the left Figure \ref{subfig:Krec} shows the case of Figure 5g while on the right Figure \ref{subfig:Kadv} shows the case of Figure 5h.}
\label{fig:anglesomparison}       
\end{figure*}

\subsection{Equivalent Contact Angle}\label{sec:res2}

We here analyze the correlation between the evolution of the contact angle and the evolution of the interface dynamics. To this end, we define an equivalent contact angle as the one that would result in the same pressure drop in the case of a spherical interface. That is, this angle is defined as 

\begin{equation}
    \theta_{D,m}(t)=\arccos{\frac{K_{exp}(t)}{R}}
    \label{eq:thetaequivalente}\,,
\end{equation} where $K_{exp}$ is evaluated from equation \ref{eq:Kterm} on the interface shape, obtained via the regression of equation \eqref{eq:thetaequivalente} (interface examples in Figures 5c-d).

This equivalent contact angle can be seen as the largest possible macroscopic contact angle that produces an equivalent impact on the interface dynamics. If the interface curvature is enslaved to the evolution of the contact angle, one would expect a correlation between $\theta_D(t)$ and $\theta_{D,m}(t)$. However, this correlation was not found in any of the experiments. Figure \ref{fig:anglesomparison} shows the evolution of $\theta_{D,m}(t)$ versus the measured $\theta_D$ (reported in Figure 5g and 5h) for both the advancing (left) and the receding (right) experiments. The marker colors are linked to the time $t$ allows following the trajectories of the points in this plane; these figures should be analyzed together with the interface oscillation plots in figure 5e and 5f.

These figures shows that the equivalent contact angle is much larger than the actual one, as one might expect from the short length scale of the meniscii observed in Figure 5. In static conditions, the equivalent contact angle is $\theta_{S,m}\approx32^{\circ}$ while the maximum advancing and minimal receeding values correspond respectively to $\theta_{A,m}=40^{\circ}$ and $\theta_{R,m}=29^{\circ}$. For the purposes of this one, we note that no clear relation appears between the two angles, neither in receding nor in advancing conditions: as these quantities evolves independently, we conclude that interface shape is mostly governed by forces that are acting far from the wall, where the influence of the contact angle appears to be negligible.

This observation is of course only valid for the specific conditions analyzed in this work, and one might argue that this is due to the large tube radious $R^*=R/l_c$ in relation to the liquid's capillary lenght. We address this concern in the next section.

\subsection{A note on the force balance}\label{sec:res3}

We are interested in how the relative contribution of the four terms in \eqref{eq:utubenondimensional} change as a function of $R^*$ and $L^*$. To this end, we solve this equation for a wide range of $R^*$ and $L^*$. First, however, the solution of this equation requires a formulation for the capillary term and for the unknown coefficient $C_f$.

As we are solely interested in the orders of magnitude, we simplify the treatment of the capillary term and replace $\left(K_A^\ast(t^\ast)-K_B^\ast(t^\ast)\right)$ with $R^*(cos(\theta_{R,m}(t^\ast))-cos(\theta_{A,m}(t^\ast))) sign(\dot{\bar{h}}^\ast(t^\ast))$ as done also in a similar work by \citet{dollet2020transition}. This shifts the problem of providing an interface shape to that of providing an equivalent contact angle law. We consider the simplest possible one: a constant value for both the advancing and the receding contact angles. We take $\theta_{A,m}=40^o$ and $\theta_{A,m}=29^o$, that is the largest and the smallest values observed in Figure \ref{fig:anglesomparison}. This undoubtedly overestimates the capillary forces' role in the liquid column's dynamics. Concerning the term $C_f$, we fit it to the data using an optimization problem and obtained $C_f=25$. This value is similar to the one received by \citet{dollet2020transition}, who considered a similar geometry. 

\begin{figure*}[h!]
\begin{subfigure}[t]{0.49\textwidth}
        \centering
    \includegraphics[width=8.5cm,clip]{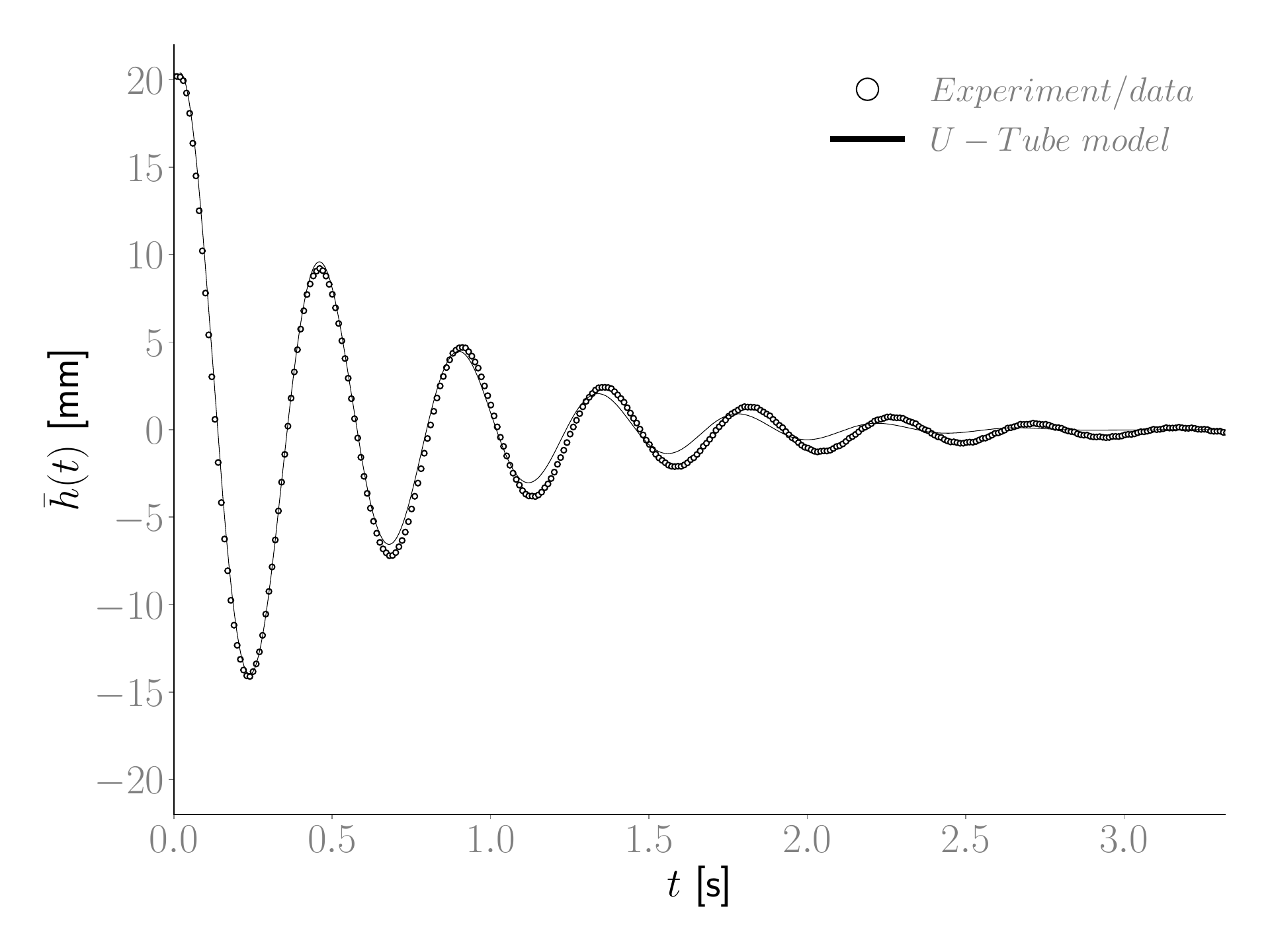}
    \caption{}
    \label{fig:experimentsimulation}
\end{subfigure}
\begin{subfigure}[t]{0.49\textwidth}
        \centering
    \includegraphics[width=8.5cm,clip]{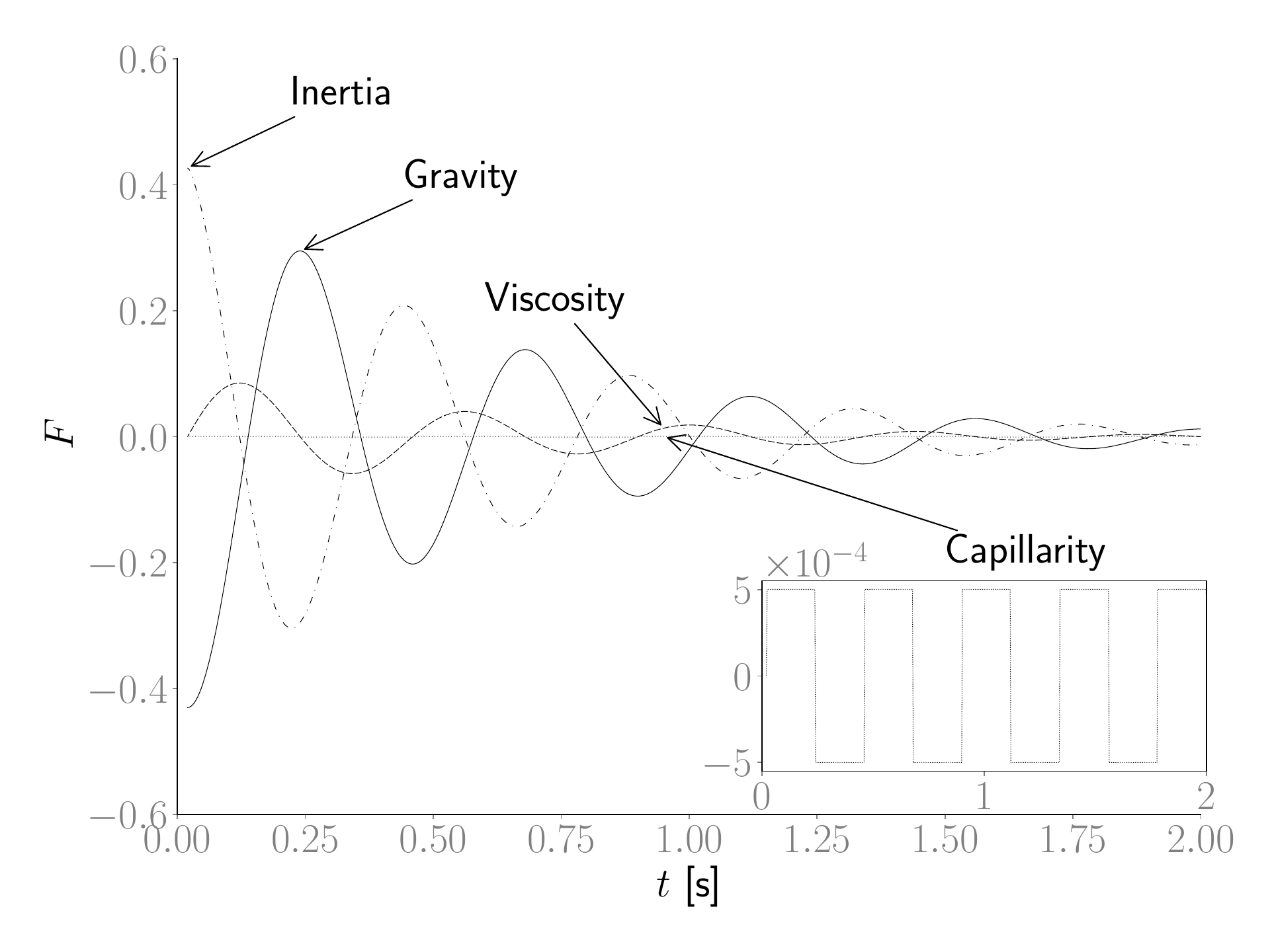}
    \caption{}
    \label{fig:experimentforces}
\end{subfigure}
\label{fig:experimentsimluationcomp}
\caption{On the left Figure \ref{fig:experimentsimulation} shows the comparison of the experiment with the interface prediction obtained solving the 1D-ODE model Equation \ref{eq:utubenondimensional}. On the right, Figure \ref{fig:experimentforces} shows the evolution of the dimensionless terms of Equation \ref{eq:utubenondimensional}.}
\end{figure*}

Figure \ref{fig:experimentsimulation} compares experimental data and model prediction with the abovementioned parameters. The predicted interface position is in good agreement with the experimental one, especially for $t<1$s. Figure \ref{fig:experimentforces} shows the history of the different terms of equation \ref{eq:utubenondimensional}. Despite the overestimation in the oversimplified step-like contact angle law (see zoom in Figure \ref{fig:experimentforces}), the capillary term is orders of magnitude lower than the others. The viscous damping is the main contribution to slowing down the interface oscillations. 

We conclude this note on the force balance using the previous closure for an extensive range of $R^*$ and $L^*$ while keeping the same oversimplified law for the contact angle. To ensure consistency with the asymptotic limits $R^*\rightarrow \infty$ and $R^*\rightarrow 0$, we introduced a smooth step-like function, such that the equivalent contact angle equals the actual ones at $R^*\ll1$ and $\approx 90^o$ at $R^*\gg1$. Adding that the resulting laws must comply with the values observed for our experiments at $R^*=3.5$ provides the following

\begin{equation}
    \theta_{A/R,m}(R^*)=\frac{\pi/2}{1+e^{-\zeta_{A/R}(R^*-R^*_{A/R,0})}}
    \label{eq:sigmoid}
\end{equation} where the parameters $\zeta, R^*_0$ are provided in table \ref{tab:shapeParams}. These were constrained by imposing $\theta_{A,m}(R^*=3.5)=40^\circ$, $\theta_{A,m}(R^*<1)\approx 25^\circ$ and $\theta_{A,m}(R^*>8)\approx 90^\circ$ to identify $\zeta_A, R^*_{A,0}$ and $\theta_{R,m}(R^*=3.5)=29^\circ$, $\theta_{A,m}(R^*<1)\approx 0^\circ$ and $\theta_{A,m}(R^*>8)\approx 90^\circ$ to identify $\zeta_R, R^*_{R,0}$.

\renewcommand{\arraystretch}{1.1}
\begin{table}[h]
\centering
	\caption{\label{tab:shapeParams} $\theta_{A,m}(R^*)$ and $\theta_{R,m}(R^*)$ shape parameters.}
		\begin{tabular}{cccc}
			\hline
			$\zeta_A$ & $R^*_{A,0}$ & $\zeta_R$ & $R^*_{R,0}$   \\
            1.49 & 3.65 & 1.24 & 4.10 \\
		\end{tabular}
\end{table}

\begin{figure*}[h]
\begin{subfigure}[t]{0.49\textwidth}
        \centering
    \includegraphics[width=9cm,trim={6cm 2cm 6cm 2cm},clip]{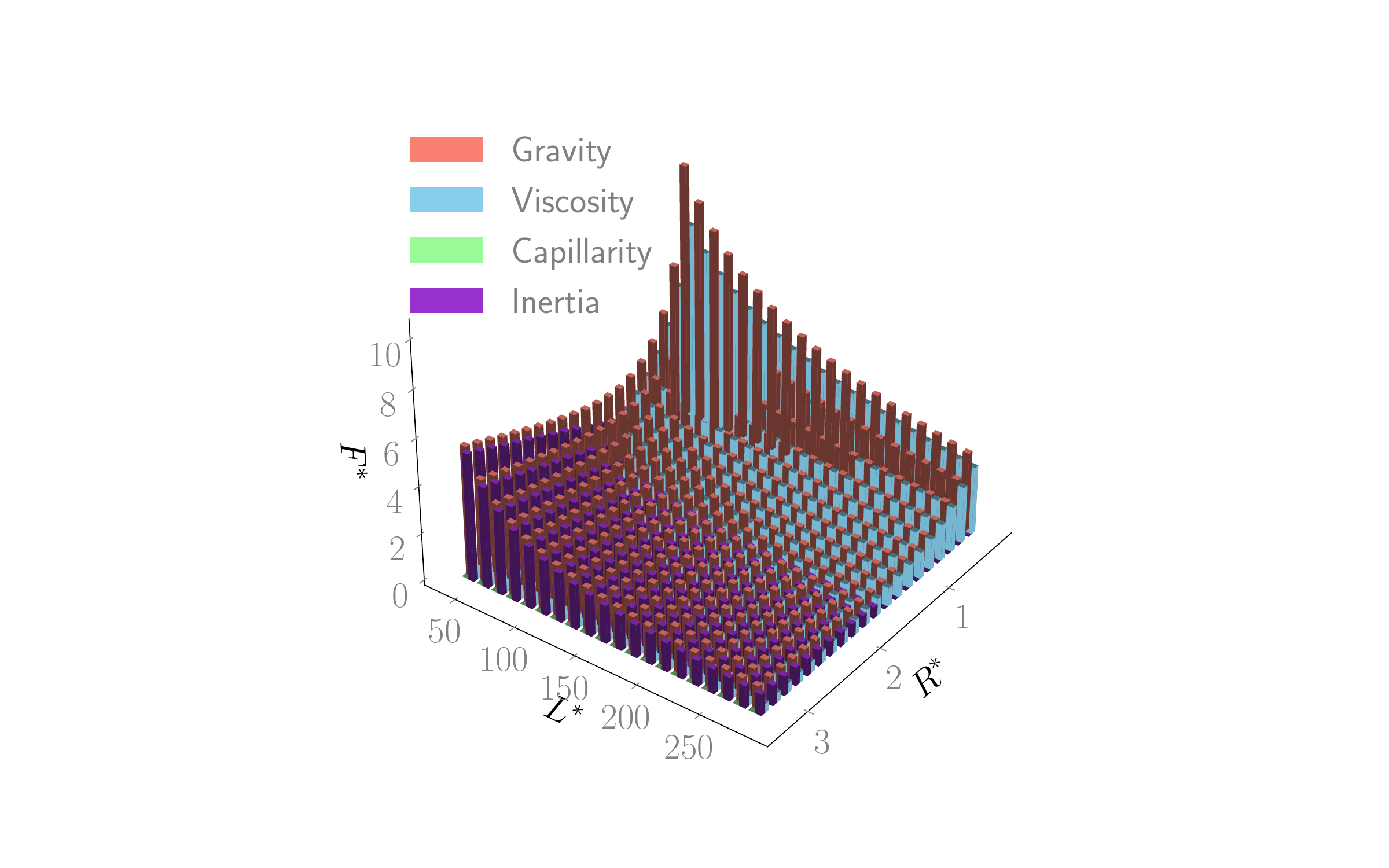}
    \caption{}
    \label{fig:barchart}
\end{subfigure}
\begin{subfigure}[t]{0.49\textwidth}
        \centering
    \includegraphics[width=9cm,clip]{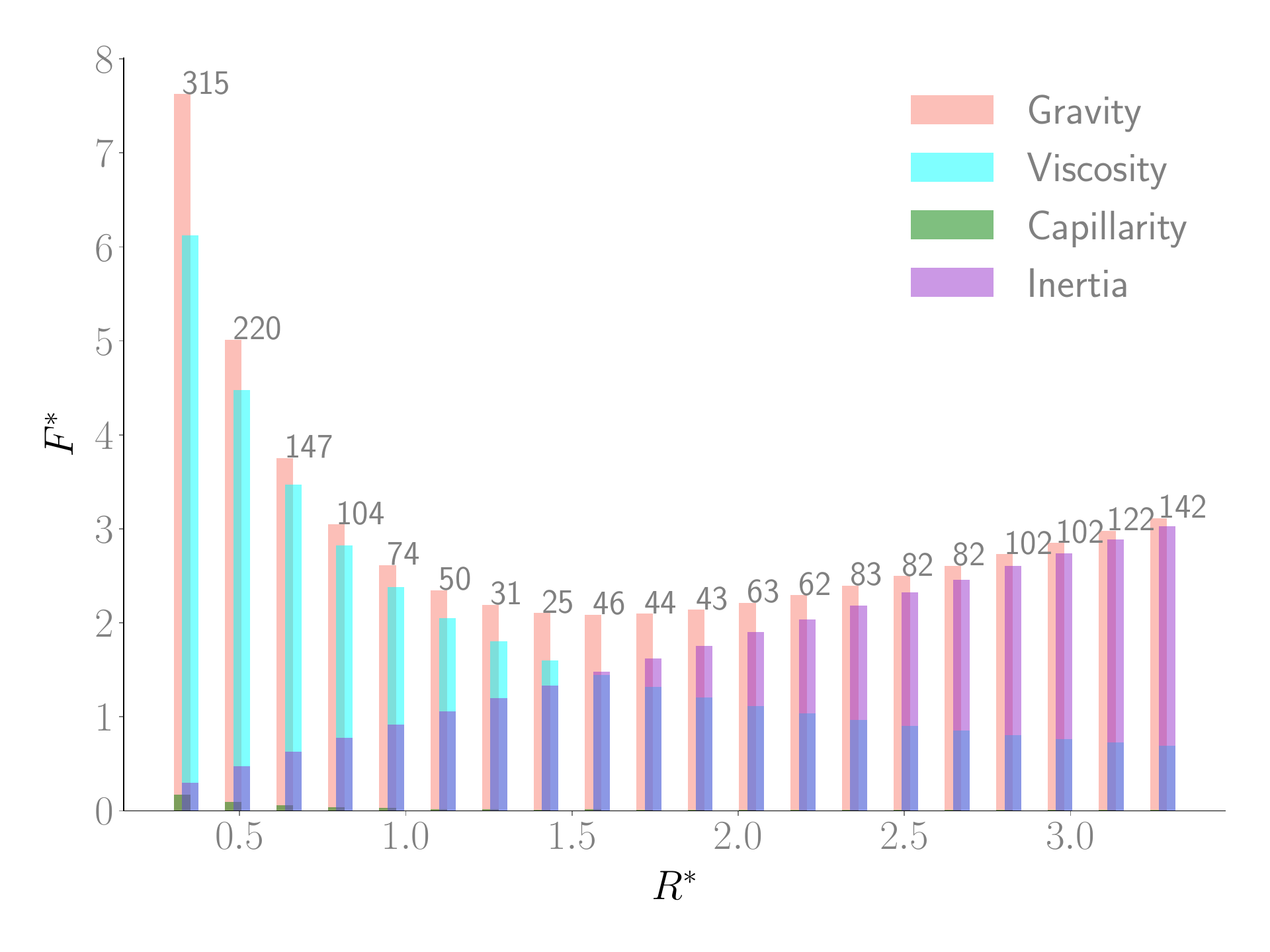}
    \caption{}
    \label{fig:bartchartforces}
\end{subfigure}
\label{fig:experimentsimluationall}
\caption{Figure \ref{fig:barchart} shows the square of the l2 norm of each force term in the model equation \ref{eq:utubenondimensional}. The plot shows that the capillary pressure drop at the interface has a minor impact also at small-size experiments because of the rising contribution of viscous forces. Figure \ref{fig:bartchartforces} shows a section of the chart \ref{fig:barchart} for $L^*=94$, together with the dimensionless duration of the virtual experiment. }
\end{figure*}

We thus simulate a total of 400 experiments, varying $R^*, L^*$ in the range $R^* \in [0.4,3.5]$ and $L^* \in [10,280]$, and considering the same dimensionless initial position $\overline{h}^*(0)=20$. For each of these, we define the experiment duration as the time for which we have simultaneously $|\bar{h}(t_x)|<0.1|\bar{h}^*(0)|$ and $|\dot{\bar{h}}^*(t_x)|<0.1(l_c g)^{1/2}$. 
Concerning the role of the viscous dissipation, we keep the same value of $C_f=25$ for all simulations. This might result in the incorrect prediction of the contribution of viscous forces for the smallest tubes. However, considering more complex correlations for the pressure drop in curved tubes (see for example \citet{ghobadi2016review}) reveals that the error can be of the order of $\sqrt{R/R_2}$ where $R_2$ is the radius of curvature of the bend. Keeping the focus on the relative order of magnitude of the different forces, this does not significantly change the results on the unexpectedly minor role of surface tension. The figure \ref{fig:barchart} shows the contribution of forces over the full range of simulated lengths and radii while the figure \ref{fig:bartchartforces} focuses on a case with $L^*=94$.

Both figures show that viscosity and gravity dominate the balance at the limit of small $R^*$ while the contribution of inertial forces rises linearly with $R^*$ and quadratically with $L^*$. The figure \ref{fig:bartchartforces} also shows the dimensionless duration of the experiment for each case. A minimum occurs at about $R^*=1$. This is the critically damped condition where the interface reaches equilibrium without oscillation. Oscillations are produced at larger radii while over-damping (and nearly first-order behavior) is observed at lower radii. The capillary contribution remains negligible in the whole investigated range. In other words, as long as these experiments are carried out in normal gravity conditions, the role of surface tension in the dynamics of the U-tube experiments is negligible for any suitable combination of parameters $R^*$ and $L^*$.

\captionsetup[subfigure]{skip=-0.8\textwidth,margin=0.2\textwidth}

\section{Conclusions}\label{sec:conclusions}

This work experimentally investigated the dynamics of a moving contact line for liquid nitrogen in a quasi-capillary U-tube (with $R^*=R/l_c\approx3.5$)  in cryogenic conditions. The experimental setup allowed for visualizing the gas-liquid interface during its motion, while image processing techniques and regression with dynamic interface models allowed for accurate dynamic contact angle detection. The contact angle evolution was compared with an unsteady generalization of Tanner-Voinov-Hoffman and the simpler Davis-Hocking linear relationship. The second proved to be valid in advancing conditions in both the case of dry and pre-wet surfaces. This result aligns with previous studies on sessile droplets on vibrating substrates. In receding conditions, the contact angle appeared less correlated with the Capillary number but close to the static contact angle. 

We analyzed the link between the contact angle and the interface evolution using a macroscopic equivalent contact angle, defined as the angle that would make a spherical interface have the same capillary pressure drop as the actual interface. The equivalent contact angle and the actual contact angle were shown to be uncorrelated, suggesting that the interface motion is independent of the wetting dynamics {when $R^*>>1$}. Finally, we analyzed the force balance governing the motion of the liquid column over a wide range of tube diameters and lengths. The results show that capillary forces play a minor role in the U-tube experiments: inertia dominates in large tubes while viscosity and gravity dominate in the small ones. Future work will consider a modification of the experiments presented here, with the aim of increasing the sensitivity of the interface motion to the capillary pressure. This will allow to analyze better the impact that dynamic wetting can have on the dynamics of a gas-liquid interface. To this end, experiments will be carried out in microgravity conditions.

\begin{acknowledgments}
	The authors thanks Mathieu Delsipee for his support and contribution in the preparation of the experimental set up. D. Fiorini is supported by Fonds Wetenschappelijk Onderzoek (FWO), Project number 1S96120N and the work was supported by the ESA Contract No. 4000129315/19/NL/MG. 
\end{acknowledgments}



%

\end{document}